\documentclass[twocolumn]{aastex63}

\usepackage{xspace}
\usepackage{amsmath}

\defcitealias{Poggianti2017}{Paper I}
\defcitealias{Bellhouse2017}{Paper II}
\defcitealias{Fritz2017}{Paper III}
\defcitealias{Gullieuszik2017}{Paper IV}
\defcitealias{Moretti2018}{Paper V}
\defcitealias{Vulcani2018}{Paper VII}
\defcitealias{Vulcani2017c}{Paper VIII}
\defcitealias{Jaffe2018}{Paper IX}
\defcitealias{Vulcani2018b}{Paper XII}
\defcitealias{Poggianti2019}{Paper XIII}
\defcitealias{Vulcani2018c}{Paper XIV}
\defcitealias{Vulcani2019b}{Paper XX}
\defcitealias{Poggianti2019b}{Paper XXIII}

\newcommand{\Ha}{H$\alpha$\xspace}
\newcommand{\Hb}{H$\beta$\xspace}

\newcommand{\kms}{$\rm km \, s^{-1}$\xspace}

\newcommand{\sinopsis}{{\sc sinopsis}\xspace}
\newcommand{\ma}{$\rm M_\ast$\xspace}
\newcommand{\ms}{$\rm M_\odot$\xspace}

\submitjournal{ApJ}

\shorttitle{Recently quenched galaxies in the GASP sample}
\shortauthors{B. Vulcani et al.}

\begin{document}

\title{GASP XXIV. The  history of abruptly quenched galaxies in clusters}

\correspondingauthor{Benedetta Vulcani}
\email{benedetta.vulcani@inaf.it}

\author[0000-0003-0980-1499]{Benedetta Vulcani}
\affiliation{INAF- Osservatorio astronomico di Padova, Vicolo Osservatorio 5, 35122 Padova, Italy}

\author{Jacopo Fritz}
\affiliation{Instituto de Radioastronom\'ia y Astrof\'isica, UNAM, Campus Morelia, A.P. 3-72, C.P. 58089, Mexico}

\author{Bianca M. Poggianti}
\affiliation{INAF- Osservatorio astronomico di Padova, Vicolo Osservatorio 5, 35122 Padova, Italy}

\author{Daniela Bettoni}
\affiliation{INAF- Osservatorio astronomico di Padova, Vicolo Osservatorio 5, 35122 Padova, Italy}

\author{Andrea Franchetto}
\affiliation{Dipartimento di Fisica \& Astronomia ``Galileo Galilei'', Universit\`a di Padova, vicolo dell' Osservatorio 3, 35122, Padova, Italy}
\affiliation{INAF- Osservatorio astronomico di Padova, Vicolo Osservatorio 5, 35122 Padova, Italy}

\author{Alessia Moretti}
\affiliation{INAF- Osservatorio astronomico di Padova, Vicolo Osservatorio 5, 35122 Padova, Italy}

\author{Marco Gullieuszik }
\affiliation{INAF- Osservatorio astronomico di Padova, Vicolo Osservatorio 5, 35122 Padova, Italy}

\author{Yara Jaff\'e }
\affiliation{Instituto de F\'isica y Astronom\'ia, Universidad de Valpara\'iso, Avda. Gran Breta\~na 1111 Valpara\'iso, Chile}

\author{Andrea Biviano}
\affiliation{INAF-Osservatorio Astronomico di Trieste, via G.B. Tiepolo 11, 34143 Trieste, Italy }
\affiliation{IFPU-Institute for Fundamental Physics of the Universe, via Beirut 2, 34014 Trieste, Italy}

\author{Mario Radovich}
\affiliation{INAF- Osservatorio astronomico di Padova, Vicolo Osservatorio 5, 35122 Padova, Italy}

\author{Matilde Mingozzi}
\affiliation{INAF- Osservatorio astronomico di Padova, Vicolo Osservatorio 5, 35122 Padova, Italy}




\begin{abstract}
The study of cluster  post starburst galaxies gives useful insights on the physical processes quenching the star formation in the most massive environments. Exploiting the MUSE data of the GAs Stripping Phenomena in galaxies (GASP) project, we characterise the quenching history of 8 local cluster galaxies that were selected for not showing emission lines in their fiber spectra. We inspect the integrated colors, and the \Hb rest frame equivalent widths (EW),  star formation histories (SFHs) and  luminosity-weighted age (LWA) maps finding no signs of current star formation throughout the disks of these early-spiral/S0 galaxies. All of them have been passive for at least 20 Myr,  but their SF declined on different timescales. 
{  In most of them  the outskirts reached undetectable SFRs before the inner regions (``outside-in quenching'')}. Our sample includes three post-starforming galaxies, two passive galaxies and three galaxies with intermediate properties. The first population shows blue colors, deep \Hb in absorption (EW$>>$2.8 \AA{}), young ages (8.8$\rm <\log(LWA [yr])<$9.2). Two of these galaxies show signs of a central SF enhancement before  quenching. Passive galaxies have instead red colors, EW(\Hb)$<$2.8 \AA{}, ages in the range 9.2$<\rm \log(LWA [yr])<$10. Finally, the other galaxies are most likely in transition between a post starforming and passive phase, as they quenched in an intermediate epoch and have not lost all the star forming features yet.
The outside-in quenching, the  morphology and kinematics of the stellar component, along with the position of these galaxies within massive clusters ($\sigma_{cl}$= 550-950 \kms) point to a scenario in which ram pressure stripping has  removed the gas, leading  to quenching. Only the three most massive galaxies might alternatively have entered the clusters already quenched.  These galaxies are therefore at the final stage of the rapid evolution galaxies undergo when they  enter the cluster environment. 
\end{abstract}



\section{Introduction}

During their life, galaxies move from the star forming blue cloud to the quiescent red sequence through a number of different pathways \citep{Barro2014, Schawinski2014, Vulcani2015}. The cessation of the star formation can take place on different time scales, strongly dependent on a galaxy's growth history \citep{Martin2007} and on a galaxy environment \citep{Balogh2004}.

Galaxy clusters are relatively hostile environments where the impact of environmental quenching should reveal itself most prominently. Many physical mechanisms may act in clusters to both trigger and truncate star formation in infalling galaxies \citep[see][for a review]{Boselli2006}. The processes can be divided into two categories: interactions between the galaxy gas  and the hot ($10^7-10^8$ K), rarefied ($10^{-3}$ particles cm$^{-3}$) intracluster medium (ICM), and   gravitational interactions either between the galaxy with other cluster members or with the cluster's gravitational potential.

Ram pressure and viscous stripping \citep{Gunn1972, Nulsen1982} are hydrodynamical interactions that fall into the first category. They can easily remove the hot gas halo reservoir, thereby leading to a gradual decline in star formation \citep[strangulation][]{Larson1980, Bekki2002, Bekki2009}, almost without affecting the structure of the old stellar population.
Strong ram pressure stripping can also remove the cold disk gas that fuels star formation, leading to a possible temporal enhancement of the star formation \citep[e.g.][]{Vulcani2018_L} and to its quenching  on short timescales \citep{RoedigerBruggen2006, Bekki2014, Boselli2014, Lee2017}.   Numerical simulations have shown that the truncation of star formation (i.e. the quenching) after ram pressure stripping can proceed outside-in \citep[e.g.][]{Kronberger2008,Bekki2009, Bekki2014}{ , that is galaxy outskirts reach undetectable star formation levels before the inner regions.}
Gravitational interactions include tides due to the cluster potential, other galaxies, or the combined effect \citep[harassment,][]{Moore1996}. These processes can disrupt both the distribution of old stars and the gas in a cluster galaxy, entailing transformations in the morphological, kinematical, star formation, and active galactic nucleus (AGN) properties of cluster galaxies \citep{Byrd1990, Bekki1999}. Galaxy-galaxy mergers are less frequent in the cluster cores because of the high relative velocities of the galaxies \citep{Ghigna1998}. Nonetheless, almost half of the galaxies in massive clusters are accreted through smaller, group-scale halos where mergers are expected to be efficient \citep{McGee2009, Haines2018, DeLucia2012}.

Each of the aforementioned processes should have a different impact 
on the star formation of recently accreted galaxies, and to act on different timescales to shut off star formation.

Post starburst (PSB) galaxies are a peculiar class of objects caught in the midst of a rapid transition from star forming to quiescent.  PSBs are currently quenched, as indicated by their lack of significant nebular emission lines.  Their spectra are characterised by  strong Balmer absorption lines that reveal a substantial population of A stars, indication that these galaxies have experienced either a ``normal'' star formation activity, or (in the cases with the strongest lines) a burst of star formation sometime in the past 1-1.5 Gyr \citep{dressler83, couch87,poggianti99}. The spectra also show signs of the presence of K-giant (older) stars, 
that dominate early-type (E) galaxy spectra. 
Such decomposition is at the origin of the names ``k+a/a+k'', and ``E+A'', often used to describe PSB galaxies. 
Given the known lifetime of  A-type stars, the evolution of this population can be used as a quenching clock. 

In the literature, the term PSB is generally used not only to refer to galaxies that indeed had a burst before the recent quenching, but also to indicate both galaxies that simply had a rapid truncation of the star formation, without having a burst and galaxies that had an abrupt reduction of the star formation, but that still show some ongoing activity \citep[e.g.,][]{Alatalo2016}. To describe the second and third subpopulations, terms like post star-forming and low-level star forming galaxies would be more appropriate.  To avoid confusion, we will use the term PSB when referring to literature results where the selection is ambiguous, $a+k/k+a$ when discussing our own results. 

Some PSBs are found in the ``green valley'' \citep[e.g.][]{Caldwell1996, zabludoff96, Norton2001, Pracy2009, Wong2012, Zwaan2013, Wu2014, Yesuf2014, Vulcani2015, Pattarakijwanich2016, Paccagnella2017, Paccagnella2019} of the optical color-magnitude diagram, indicating stellar populations will redden and evolve passively onto the red sequence, others are part of the blue cloud or, in the most evolved cases, even the red sequence. PSB morphologies \citep{poggianti99,Yang2004, Yang2008, Poggianti2009b} and spatially resolved kinematics \citep{Norton2001, Swinbank2012} are also consistent with late-type galaxies evolving into early-type galaxies.

Many evolutionary channels to quench PSB galaxies have been proposed, to account for the diversity of galaxy properties \citep[e.g.,][]{Pawlik2019}.  Two broad scenarios can explain the formation of PSBs in different environments.
The presence of PSBs in clusters, where \cite{DresslerGunn1982} first observed them, is commonly believed to be due to ram-pressure stripping. Gas-rich star-forming galaxies fall into clusters and their gas is removed by the interaction with the ICM, suddenly quenching their star formation \citep{DresslerGunn1982, couch87, DresslerGunn1992, poggianti99, baloghnavarromorris00, Tran2003, Tran2004, Tran2007,  Poggianti2009b,  Paccagnella2017}. In this traditional scenario, as stellar morphologies are not affected by ram pressure stripping, the resulting PSB galaxies still resemble disc galaxies,
at least until the effects of the past star formation fade away leaving an S0 galaxy \citep{Bekki2002}. This interpretation holds also for PSB galaxies in massive groups, where ram pressure is still effective \citep[e.g.,][]{Poggianti2009b, Paccagnella2019}. 
In some cases cluster mechanisms may quench galaxies without triggering any significant burst of star formation \citep{Socolovsky2018}.

In the field PSB galaxies show disturbed kinematics and tidal features indicative of violent relaxation due to major, late-stage mergers \citep{zabludoff96, Yang2004, Blake2004, Tran2004, Goto2005, Yang2008,  Pracy2009, French2016}. Merger induced supernova or AGN feedback can also aid the expulsion of the gas that becomes too hot to collapse further or is expelled altogether \citep{Sanders1988, Hopkins2006}.
This merger scenario is supported also by the detection of Balmer gradients and  young stellar populations  centrally concentrated with respect to the old population (\citealt{Norton2001, Pracy2012, Pracy2013} - but see \citealt{Chilingarian2009, Yagi2006, Pracy2009, Swinbank2012} who failed to recover such trends). 
Moreover, field PSBs show a range of angular momentum properties, consistent with a variety of possible merger histories \citep{Pracy2009, Pracy2013, Swinbank2012}. 

Intriguingly, field PSBs have been found to hold large molecular \citep{French2015, Rowlands2015} and cold \citep{Zwaan2013} gas reservoirs, ruling out complete gas consumption, expulsion or starvation as the primary mechanism that ends the star formation. Significant gas reservoirs in PSB galaxies have also been predicted by simulations \citep{Davis2019}, who found that a variety of gas consumption/loss processes are responsible for the rapid evolution of this population, including mergers and environmental effects, while active galactic nuclei play only a secondary role. 
No observations of molecular gas of cluster PSBs are available to date.

In this context, the  advent of  integral field spectroscopy (IFS) is extremely useful to characterise the spatial distribution of the stellar populations and obtain information about the mechanism responsible of the PSB features. 
First studies focused on few tens of PSBs 
and in most cases only the central regions of the galaxies have been observed \citep{Chilingarian2009, 
Pracy2009, 
Swinbank2012, 
Pracy2012, 
 Pracy2013}.

With the advent of large IFU spectroscopic surveys, more detailed analysis have been performed. \cite{Chen2019} analyzed 360 galaxies with either central PSB regions, or with off-center ring-like PSB regions in the Mapping Nearby Galaxies at APO \citep[MaNGA,][]{Bundy2015} survey. They showed that these galaxies are not simply different evolutionary stages of the same event, rather the former are caused by a significant disruptive event that produced a rapid decline of star formation in the central region, while the latter are caused by disruption of gas fuelling to the outer regions.

Other IFS studies did not targeted specifically PSBs, but managed to characterise PSB features in star forming galaxies \citep[e.g.,][]{Poggianti2017, Gullieuszik2017, Poggianti2019b}. \cite{Roche2015} investigated a merging system in the Calar Alto Legacy Integral Field Area  \citep[CALIFA,][]{sanchez06} survey and found  that much of the galaxy, especially the outer tidal arms, has a PSB spectrum, evidence of a
more extensive recent  episode of star-formation, triggered by the previous
perigalacticon passage. \cite{Rowlands2018} have used  data from MaNGA to derive Star Formation Histories (SFHs)  of different galaxy subpopulations finding that PSB regions are more common outside of the galaxy center, are preferentially found in asymmetric galaxies, and have lower gas-phase metallicity than other regions, consistent with interactions triggering starbursts and driving low-metallicity gas into regions at $<$1.5 effective radii. \cite{Owers2019} used the Sydney-AAO Multiobject Integral field spectrograph  Galaxy Survey \citep[SAMI,][]{Bryant2015} data to characterise the rare ($\sim$2\% of galaxies with  $\mathrm{log}({M}_{* }/{M}_{\odot }) >$ 10) population of galaxies showing  PSB features  in  more than 10\% of their spectra. 
These galaxies are more frequent in clusters than in the low-density environments, representing 15\% and 2\% of non-passive  galaxies, respectively. In clusters, PSB regions are confined to the galaxy external regions, while the centers are still star forming.  Conversely, in low density environments the  PSB signal is spread across the galaxies. 
The \cite{Owers2019} study is the only one explicitly focused on cluster galaxies.\footnote{The \citet{Pracy2009, Pracy2013} samples did include 2 and 1 cluster galaxies in their sample, but did not discuss the role of their environments on shaping their properties.} Also based on the analysis of the location of these galaxies within their clusters, they concluded that the galaxies recently entered the clusters and are currently being quenched by ram pressure stripping.

This paper presents the characterisation of the stellar properties of eight cluster galaxies targeted by the GAs Stripping Phenomena in galaxies with MUSE (GASP\footnote{\url{http://web.oapd.inaf.it/gasp/index.html}}, \citealt{Poggianti2017}, Paper I) survey,  an  ESO Large program aimed at characterizing where, how, and why gas can get removed from galaxies. These galaxies were chosen to show passive or PSB features in their fiber spectra \citep{Paccagnella2017}. Their selection was performed using the equivalent width of the H$\delta$. Thanks to the extremely high  quality of the data obtained with the integral field spectrograph Multi Unit Spectroscopic Explorer (MUSE), 
we can investigate the quenching history of these galaxies. We  note however that the MUSE wavelength coverage does not allow to observe H$\delta$ at $z\sim 0.06$, therefore our analysis will be instead based on the \Hb. 

The galaxies presented in this paper were included in the GASP survey as galaxies at the final stage of their evolution. By construction, GASP allows us to study galaxies in various stages of ram pressure stripping \citep[GASP IX]{Jaffe2018}, from pre-stripping (undisturbed galaxies, e.g. \citealt{Vulcani2018_L} - GASP XIV, \citealt{Vulcani2019b} - GASP XX), to initial stripping, peak stripping (\citealt{Bellhouse2017} - GASP II, \citealt{Bellhouse2019} - GASP XV, \citealt{Gullieuszik2017} - GASP IV, \citetalias{Poggianti2017}, \citealt{Moretti2018} - GASP V, \citealt{George2018} - GASP XI, \citealt{George2019} - GASP XVII), and post-stripping \citep[GASP III]{Fritz2017}, passive and devoid of gas \citep[e.g.,][GASP XIV]{Vulcani2018_g}.

We adopt a \cite{Chabrier2003} initial mass function (IMF) in the mass range 0.1-100 M$_{\odot}$, and the cosmological constants   $\Omega_m=0.3$, $\Omega_{\Lambda}=0.7$ and H$_0=70$ km s$^{-1}$ Mpc$^{-1}$.

\section{The sample}
The eight GASP galaxies analysed in this work were selected from the OMEGAWINGS survey \citep{Gullieuszik2015}. According to their fiber spectra (diameter 2.$^{\prime\prime}$16) obtained with an AAOmega@AAT spectroscopic campaign \citep{Moretti2017}, six of these galaxies presented $a+k/k+a$ spectra, while two had $k$ features \citep{Paccagnella2017}. 
Specifically, all have no emission lines, but the  former display a combination of signatures typical of both K- and A-type stars with strong $H\delta$ in absorption (EW($H\delta) > 3$\AA{}) - indicative of post- starburst/post-starforming galaxies whose star formation was suddenly truncated at some point during the last 0.5-1 Gyr, while the latter have weak $H\delta$ in absorption (EW($H\delta) < 3$ \AA{}) and present spectra resembling those of K-type stars, normally found in passively evolving elliptical galaxies.

Given the lack of ionised gas (see below), it is not possible to assess the presence of the AGN using the standard diagnostic diagrams \citep{Baldwin1981}. Also a search into the XMM and Chandra archives did not retrieve any bright source ($L_X>10^{42}$ erg/s) within 5$^{\prime\prime}$ from the galaxy centers. Nonetheless, in the very unlikely case that an AGN is present, it must be a very low-luminosity one. For example, Gonzalez-Martin et al. (2006, 2009) detected LINER-like emission due to AGN activity in objects with X-ray luminosity in the range $1^{38}-1^{42}$ erg/s. Therefore, it should not be the main contributor to the removal of gas \citep{Davis2019}.  

The whole  OMEGAWINGS parent sample of PSB galaxies is characterised in detail in \cite{Paccagnella2017, Paccagnella2019}.

\begin{figure*}
\centering
\includegraphics[scale=0.35]{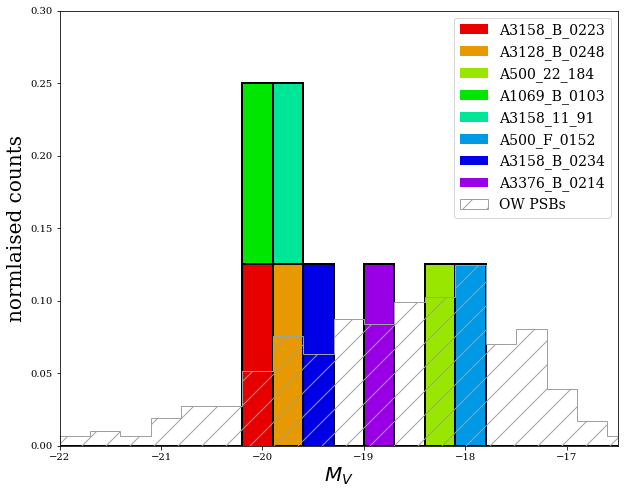}
\includegraphics[scale=0.35]{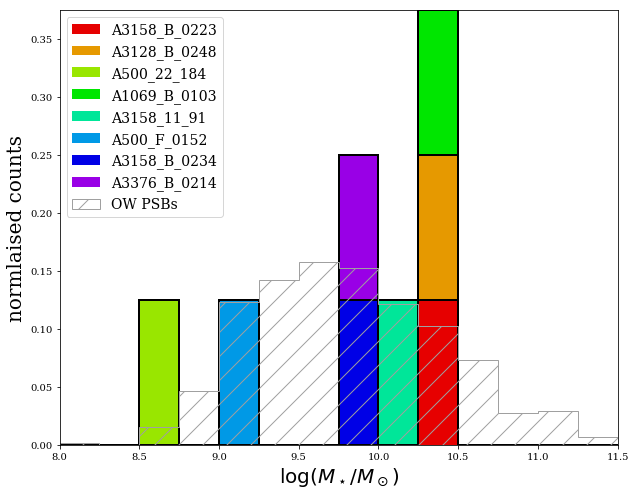}
\includegraphics[scale=0.35]{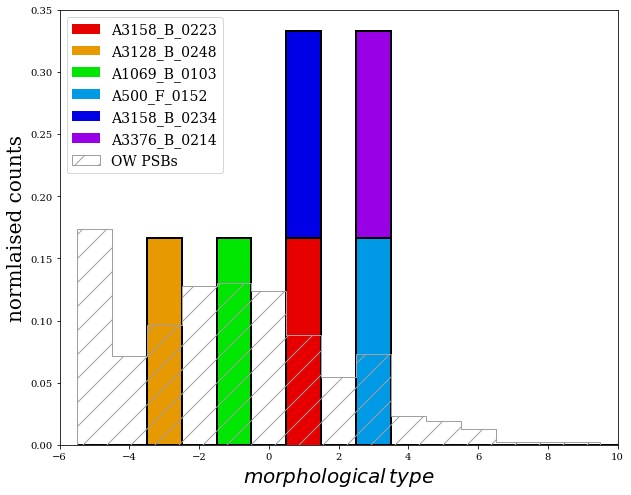}
\includegraphics[scale=0.35]{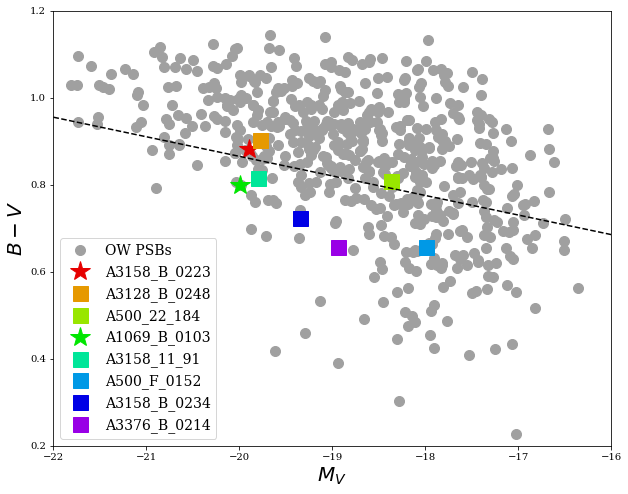}
\caption{Absolute V magnitude, stellar mass, morphological distribution and color magnitude plot of the galaxies discussed in this paper (coloured histograms and points), compared to the results presented in \citet{Paccagnella2017} (OW=OMEGAWINGS, light grey histograms and points). In the color-magnitude plot $k$ galaxies are represented by stars, $k+a/a+k$ galaxies by squares. { Galaxies are rainbow-colored by increasing EW(\Hb), as measured on the integrated spectra (see Tab.\ref{tab:EW})}. The same color scheme will be adopted throughout the paper. 
\label{fig:cf_P17} }
\end{figure*}

\begin{table*}
\caption{Galaxy name, coordinates, redshift, stellar mass, cluster, inclination, effective radius, ellipticity, position angle,  spectral type according to the fiber OmegaWINGS spectrum and morphology of the sample analysed in the paper.  \label{tab:gals}}
\centering
\begin{small}
\begin{tabular}{l|l|l|r|r|l|r|r|r|r|c| c }
  \multicolumn{1}{c|}{id} &
  \multicolumn{1}{c|}{RA} &
  \multicolumn{1}{c|}{DEC} &
  \multicolumn{1}{c|}{z} &
  \multicolumn{1}{c|}{Mass} &
  \multicolumn{1}{c|}{Cluster} &
  \multicolumn{1}{c|}{i} &
  \multicolumn{1}{c|}{$r_e$} &
  \multicolumn{1}{c|}{$\varepsilon$} &
  \multicolumn{1}{c|}{PA} &
  \multicolumn{1}{c|}{TYPE}  &
  \multicolumn{1}{c}{T$_M$}  \\
  \multicolumn{1}{c|}{} &
  \multicolumn{1}{c|}{(J2000)} &
  \multicolumn{1}{c|}{(J2000)} &
  \multicolumn{1}{c|}{} &
  \multicolumn{1}{c|}{($10^{9} \, M_\odot$)} &
  \multicolumn{1}{c|}{} &
  \multicolumn{1}{c|}{(deg)} &
  \multicolumn{1}{c|}{($^{\prime\prime}$)} &
  \multicolumn{1}{c|}{} &
  \multicolumn{1}{c|}{(deg)} &
  \multicolumn{1}{c|}{} &
  \multicolumn{1}{c}{}   \\
\hline
  A3158\_B\_0223 & 03:41:59.80 & -53:28:04.240 & 0.05621 & 23.4 & A3158 &  20 & 3.2$^{+0.1}_{-0.1}$ & 0.05$\pm$0.06 & 150  & k & S\\
  A3128\_B\_0248 & 03:29:23.42 & -52:26:02.871 & 0.05302 & 19.3 & A3128 &  67 & 1.9$^{+0.1}_{-0.1}$ & 0.59$\pm$0.02 & 24& k+a & S0\\
  A500\_22\_184 & 04:38:46.41 & -22:13:22.368 & 0.07248 & 0.34 & A500  & 67 & 1.4$^{+0.1}_{-0.1}$ & 0.60$\pm$0.03 & 169 & a+k & ?\\
  A1069\_B\_0103 & 10:39:36.49 & -08:56:34.822 & 0.06348 & 22.5 & A1069 & 70 & 1.8$^{+0.1}_{-0.2}$& 0.64$\pm$0.04 & 98 & k & S0\\ 
  A3158\_11\_91 & 03:41:16.75 & -53:24:00.580 & 0.06135 & 13.9 & A3158 &  29 & 4.6$^{+0.3}_{-0.3}$ & 0.12$\pm$0.04 & 4 & a+k & ?\\ 
  A500\_F\_0152 & 04:38:21.25 & -22:13:02.197 & 0.06951 & 1.4 & A500  & 61 & 2.6$^{+0.3}_{-0.3}$ & 0.50$\pm$0.04 & 23& a+k & S\\
  A3158\_B\_0234 & 03:42:24.68 & -53:29:25.989 & 0.06602 & 7.1 & A3158 &  50 & 2.5$^{+0.1}_{-0.1}$ & 0.35$\pm$0.05 & 126& a+k &S\\
  A3376\_B\_0214 & 06:00:43.18 & -39:56:41.641 & 0.04728 & 7.7 & A3376 &  64 & 2.8$^{+0.1}_{-0.1}$ & 0.54$\pm$0.01 & 10 & k+a & S\\

\end{tabular}
\end{small}
\end{table*}

Figure \ref{fig:cf_P17} compares the integrated properties of our galaxies to those discussed in \cite{Paccagnella2017}. The determination of such properties is deferred to Section \ref{sec:methods}. The eight galaxies span the absolute magnitude range -20$<$$\rm M_V$$<$-18 and the mass range 8.5$<$$\log$(\ma/\ms)$<$10.4. They are therefore located in the central part of the  distributions of the entire OMEGAWINGS population. { The Kolmogorov Smirnov tests detects differences only when comparing the magnitude distribution (statistic=0.99, pvalue=5e-08), while in the other cases distributions are indistinguishable (pvalue$>$0.3)}. The color magnitude diagram reveals that all the galaxies are quite close to the border between the red and blue population  \citep[$(B-V)_{rf} =-0.045\times V-0.035$,][]{Paccagnella2017},  highlighting the transitioning phase of this class of objects. More specifically, three galaxies are already red, three are still blue, two are in the green valley. It is interesting to note that one of the previously selected as $k$ galaxy is still on the blue side. The morphological analysis reveals that two galaxies (one $k$ and one $k+a$) are S0s, while the rest are early-spirals. For two galaxies the morphologies could not be determined. Morphologies were derived with MORPHOT \citep{Fasano2012}, an automatic tool designed to reproduce as closely as possible the visual classifications.

The properties of the galaxies discussed in this paper are therefore consistent with the general cluster PSB population and can be considered as representative of it. Their analysis could therefore improve our understanding of the galaxy quenching mechanisms in clusters.

\section{Methods, Observations} \label{sec:methods}

The  galaxies were observed in
service mode using the MUSE spectrograph mounted on the Very Large Telescope 
(VLT) in Paranal. 
A complete description of the survey strategy, observations, data reduction and analysis procedure is presented in \citetalias{Poggianti2017}. 

Most of the analysis presented in this paper is based on the outputs of our spectrophotometric code \sinopsis \citepalias{Fritz2017}. { This code searches the combination of Single Stellar Population (SSP) model spectra that best fits the equivalent widths of the main lines in absorption and emission and the continuum at various wavelengths, minimizing the $\chi$ = 1 using an adaptive simulated annealing algorithm \citep{Fritz2007, Fritz2011}. The star formation history is let free with no analytic priors.
\sinopsis uses a \cite{Chabrier2003} IMF with stellar masses in the 0.1-100 M$_\odot$ limits, and they cover metallicity values from Z = 0.0001 to 0.04. The metallicity of the best fit models is constant and homogeneous (i.e. all the SSPs have the same metallicity independently of age). The best fit model is searched using SSP models at three different metallicity values (sub-solar, solar and super-solar).
 Dust extinction is accounted for by adopting the Galaxy extinction curve \citep{Cardelli1989}. \sinopsis uses the latest SSP models from S. Charlot \& G. Bruzual (in prep.) based on stellar evolutionary tracks from \cite{Bressan2012} and stellar atmosphere spectra from a compilation of different authors, depending on the wavelength range, stellar luminosity, and effective temperature. \sinopsis also includes the nebular emission lines for the young (i.e., age $<$2$\times 10^7$ yr) SSPs computed with the Cloudy code \citep{Ferland2013}.} The code provides for each MUSE spaxel rest frame magnitudes,  stellar masses, luminosity-weighted and mass-weighted ages and SFHs in  twelve fine age bins.  These bins have also been combined in four logarithmically spaced age bins in such a way that the differences between the spectral characteristics of the stellar populations are maximal \citepalias{Fritz2017}. We consider as reliable only spaxels with S/N$>$3 across the entire spectrum.

Equivalent widths (EW) were measured using {\sc SPLAT},\footnote{\url{http://star-www.dur.ac.uk/~pdraper/splat/splat.html}} a publicly available graphical tool. Observed EWs are converted to rest-frame values dividing the measurements by (1 + z). We adopt the usual convention of identifying absorption lines with positive values. As explained later, EWs were measured only on  spectra integrated across portions of galaxies, to reduce the noise. 

 Total masses are obtained by summing the values of { stellar mass obtained from \sinopsis on the single spaxels} belonging to the galaxy, i.e. the region containing the spaxels whose near-\Ha continuum flux is $\sim 1 \sigma$ above the background level \citep[as in][]{Vulcani2018_L}.
 
We will also discuss the properties of the stellar component. The stellar kinematic was  derived from the analysis of the characteristics of absorption lines, using the pPXF software \citep{Cappellari2004}, which works in Voronoi binned regions of given S/N (10 in this case; see \citealt{Cappellari2003}). The value of the stellar radial velocity was further smoothed using the two-dimensional local regression techniques (LOESS) as implemented in the Python code developed by M. Cappellari.\footnote{\url{http://www-astro.physics.ox.ac.uk/~mxc/software}}

\begin{figure*}
\centering
\includegraphics[scale=0.55]{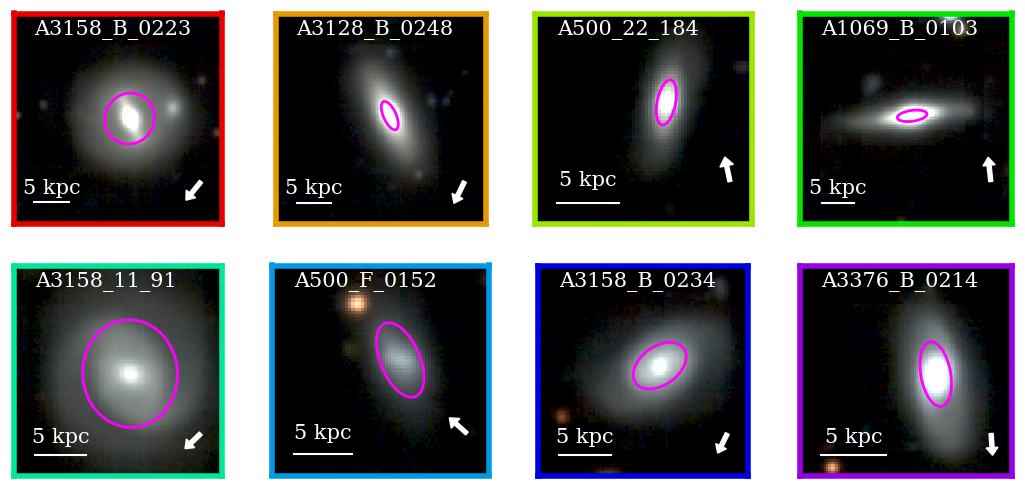}
\caption{RGB images of the galaxies used in this paper, sorted by increasing stellar mass. The reconstructed $g$, $r$, $i$  filters from the MUSE cube have been used. 
North is up, and east is left.  Magenta ellipses show the effective radius of the galaxies. The black arrows in the lower right corners show the direction of the cluster center. { Galaxies are rainbow-colored by increasing EW(\Hb), as measured on the integrated spectra (see Tab.\ref{tab:EW})}.
\label{fig:color_images} }
\end{figure*}

The structural parameters of the galaxies (effective radius $R_e$,  ellipticity $\varepsilon$, position angle $PA$ and inclination $i$) were obtained from the analysis of the images achieved from the integrated MUSE datacubes, using the Cousins I-band filter response curve, as explained in Franchetto et al. (submitted). Briefly, they were obtained using the {\sc ellipse} task \citep{Jedrzejewski1987} 
of the software IRAF. We measured the radius of an ellipse including half of the total light of the galaxies. Then, from the surface brightness profile we selected the isophotes that trace the stellar disk to measure their mean $PA$, $\varepsilon$  and corresponding errors. $i$ is derived from the apparent flattenging assuming an intrinsic axis ratio of 0.15 and will be also used to deproject galaxy properties when studying galaxy gradients (sec. 5 and 6).

Table \ref{tab:gals} presents some properties of the targets, that will be discussed in the rest of the paper. Spectral types are based on fiber spectra \citep{Paccagnella2017}. It is interesting to note that the two $k$ galaxies are also the most massive ones of the sample. In the following we will provide a new classification scheme, based on the spatially resolved properties of the galaxies. 

\begin{figure*}
\centering
\includegraphics[scale=0.25]{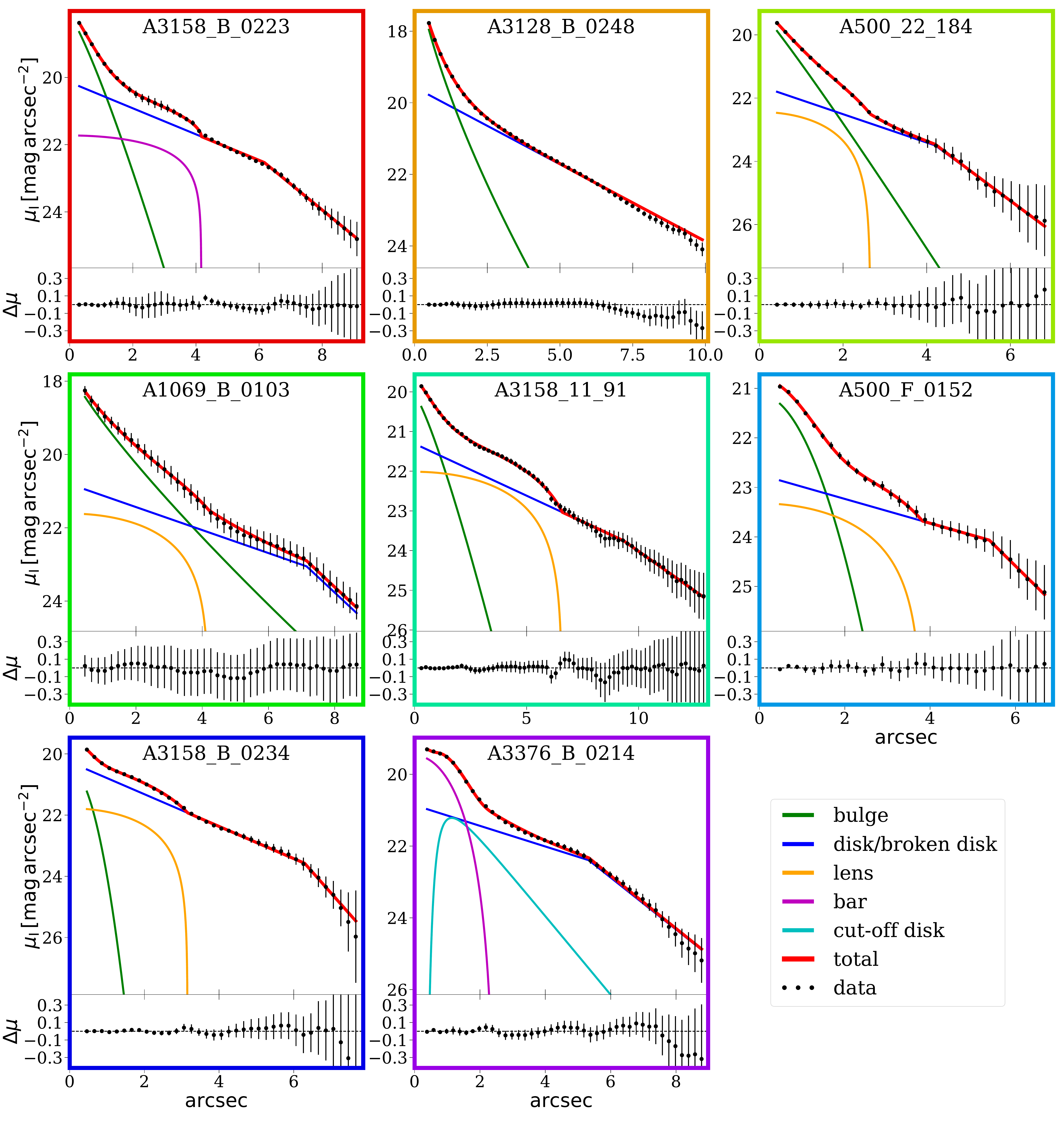}
\caption{Surface brightness profiles  and errors as derived from the I band images of the eight galaxies, sorted by increasing stellar mass. For each galaxy, the main panel shows the decomposition, the bottom panel the residual of the data with respect to the fit. Black points with errors are the data, the red lines represent the estimated surface brightness profile. The different components used to decompose the galaxies are represented with different colors, as indicated in the labels. { Galaxies are rainbow-colored by increasing EW(\Hb), as measured on the integrated spectra (see Tab.\ref{tab:EW})}
\label{fig:SBP} }
\end{figure*} 

\begin{figure*}
\centering
\includegraphics[scale=0.6]{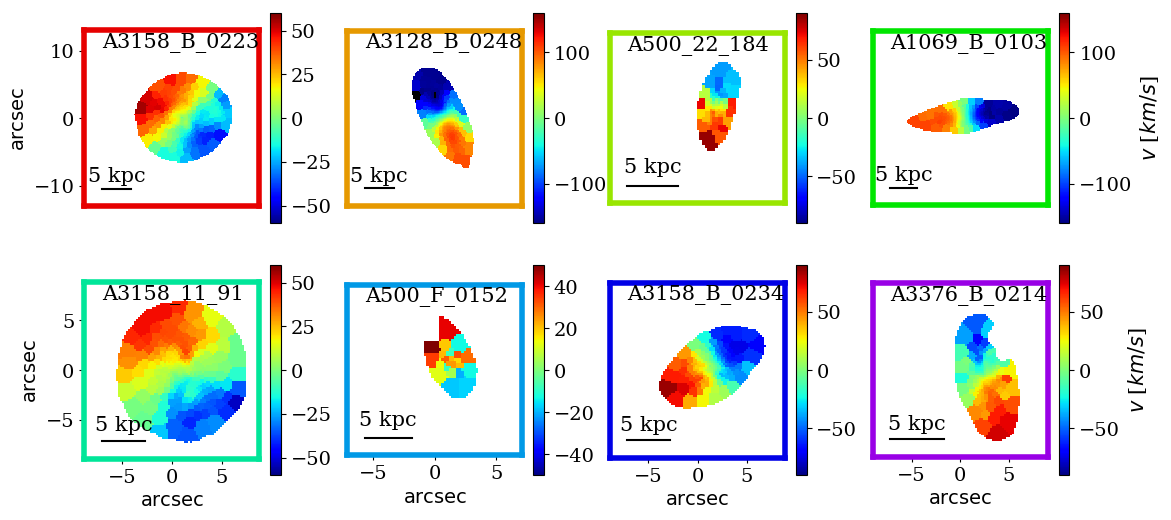}
\caption{Stellar velocity map for Voronoi bins with S/N$>$10 of the 8 galaxies. Stellar radial velocities were further smoothed using the two-dimensional local regression techniques (LOESS).
Galaxies are surrounded by squares colored following the scheme of Fig.\ref{fig:cf_P17}.
{ Galaxies are rainbow-colored by increasing EW(\Hb), as measured on the integrated spectra (see Tab.\ref{tab:EW}).}
\label{fig:vel} }
\end{figure*} 

\section{Galaxy morphologies}\label{sec:results1} 
Figure \ref{fig:color_images} shows the galaxy images based on the reconstructed $g$, $r$, $i$  filters from the MUSE cube. Each galaxy is surrounded by a color that will be used throughout the paper to identify galaxies. 
This Figure shows that  galaxies present either an early spiral or an S0 morphology, with both bulges and disks clearly distinguishable. Galaxies are also characterized by a wide range of inclinations, going from 30 to 70 degrees. They do not present evident signs of interaction with companions. The object visible on the north-west side of A3158\_B\_0223 is a background passive galaxy (z=0.064). { Since the object has no emission lines, ita presence does not affect the forthcoming results it will not be masked.} 
A3376\_B\_0214 also contains two background galaxies towards north-west. These galaxies are star forming, at a redshift of 0.7596 and they will be masked in the forthcoming analysis.   

Figure \ref{fig:SBP} shows the photometric decomposition of the galaxies, performed on the I band MUSE images. Details on the procedure adopted to obtain the parameters of the different components are given in Franchetto et al. (in prep.) and in Appendix \ref{app:SBP}. The Figure unveils that galaxies are composed by many subcomponents. All galaxies but A3376\_B\_0214 are characterised by the presence of a non-negligible bulge, whose size ranges from 1$^{\prime\prime}$ to 8$^{\prime\prime}$ at $\rm \mu_I= 25 \, mag/arcesc^2$. Only A3128\_B\_0248 can be decomposed just in terms of a bulge and a disk, while all the other galaxies present a more complex structure that includes either a bar or a lens and are characterised by a broken disk (Type II). 

This  finding might suggest that bars and other internal structures may play a significant role in triggering bursts of star formation in these galaxies or, viceversa, that during the quenching process or the SF enhancement prior to it a bar or another structure is formed.

No signs of interactions are  evident from Fig. \ref{fig:vel}, which shows the stellar velocity map of the galaxies. Galaxies present a rather undisturbed morphology and kinematics and regular rotation. Some distortions are evident in A3158\_11\_91 and A3376\_B\_0214 but these are most likely due to the presence of bars and lenses. Stellar velocity dispersions are also generally quite low ($<50$ \kms, plots not shown).

In what follow, when measuring gradients, we will not take into account the galaxy internal structure, but it will be important to keep in mind the coexistence of different components. 

\section{Equivalent width of absorption lines}
\begin{figure*}
\centering
\includegraphics[scale=0.35]{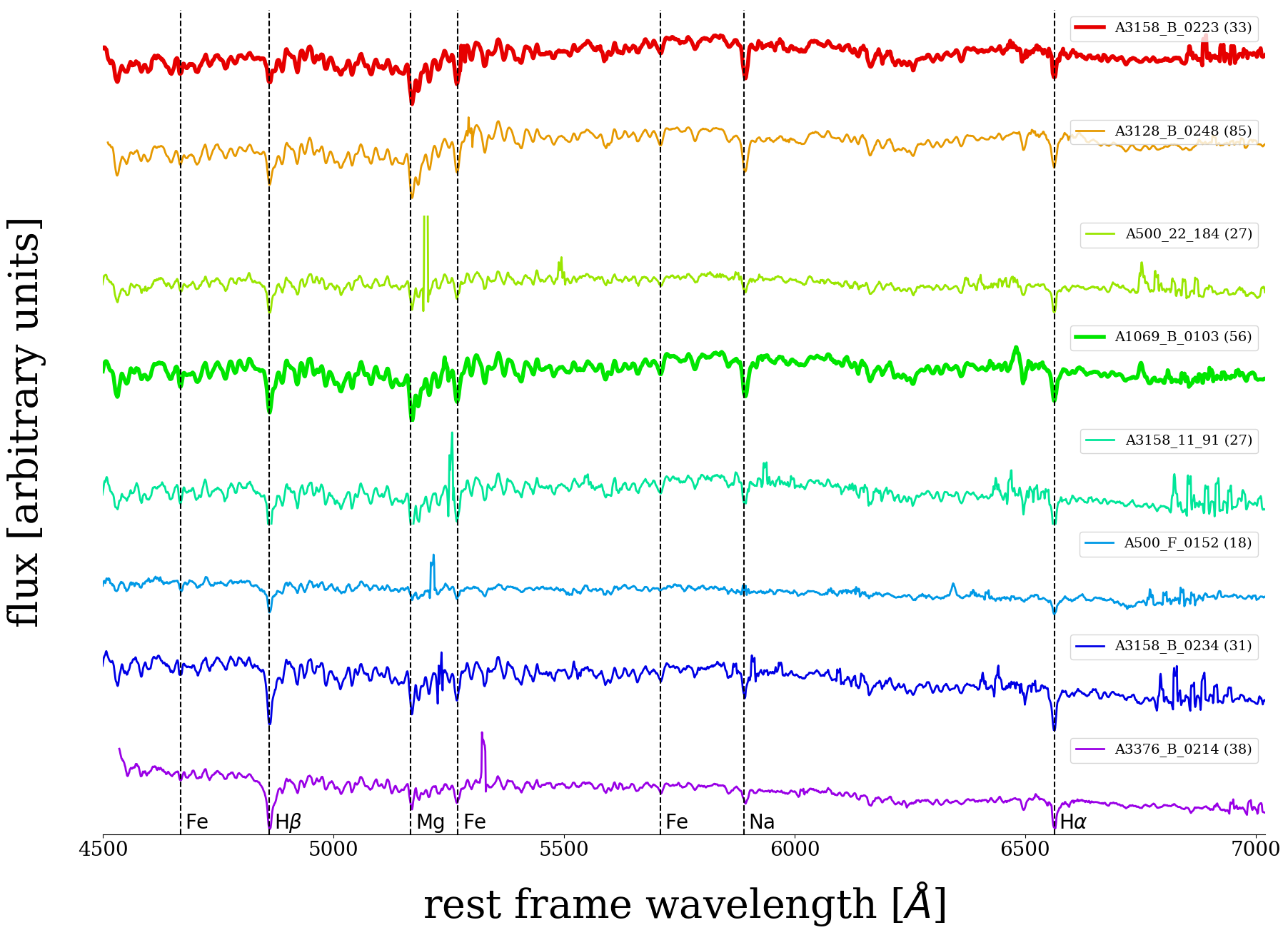}
\caption{Integrated spectrum of the galaxies in the sample. Spectra are arbitrarily shifted for display purposes. Galaxies represented by thicker lines have $k$ fiber spectra. { Numbers in parenthesis are the mean S/N ratio as measured on the entire integrated spectrum. { Galaxies are rainbow-colored by increasing EW(\Hb), as measured on the integrated spectra (see Tab.\ref{tab:EW})}}
\label{fig:spectra}  }
\end{figure*}

\begin{table}
\caption{Rest frame EW(\Hb) measured on the integrated spectra.  \label{tab:EW}}
\centering
\begin{small}
\begin{tabular}{l|c }
  \multicolumn{1}{c|}{id} &
  \multicolumn{1}{c}{EW(\Hb)} \\
  \multicolumn{1}{c|}{} &
  \multicolumn{1}{c}{(\AA{})}   \\
\hline
  A3158\_B\_0223 & 2.28$\pm$0.05 \\
  A3128\_B\_0248 & 2.65$\pm$0.06 \\ 
  A500\_22\_184 & 3.10$\pm$0.05 \\ 
  A1069\_B\_0103 & 3.12$\pm$0.04 \\
  A3158\_11\_91 & 3.12$\pm$0.04 \\ 
  A500\_F\_0152 & 3.74$\pm$0.04 \\ 
  A3158\_B\_0234 & 4.38$\pm$0.05 \\
  A3376\_B\_0214 & 5.25$\pm$0.07 \\ 
\end{tabular}
\end{small}
\end{table}

Fig. \ref{fig:spectra} shows the integrated spectra obtained from MUSE. 
Galaxies  show no emission lines, therefore ionised gas, demonstrating that there is no significant level of star formation throughout the whole galaxy, and not only in the  regions probed by the fiber spectra \citep{Paccagnella2017}.
Figure \ref{fig:spectra} also highlights how all galaxies show deep H$\beta$ in absorption, as well as a number of other lines. Unfortunately, the MUSE wavelength coverage does not allow us to sample the H$\delta$ line, upon which most of the PSB analysis in the literature relies. In what follows we will therefore characterise the \Hb instead. 
Table \ref{tab:EW} gives the values of the rest frame  EW(\Hb) for the galaxies, measured on the integrated spectrum. Values are the average of ten measurements, obtained independently shifting each time the continuum bands of few \AA{} and the \Hb band of few tenths of \AA{}. Reported uncertainties are the standard errors on the means. 

To put these numbers in context, we have measured the  EW(\Hb) of a set of model spectra. These  spectra were built  from the same set of SSP models used in \sinopsis (see \S\ref{sec:methods}), and using a solar metallicity value, assuming a two-parameters analytical 
SFH
, represented by a log-normal distribution \citep{Gladders2013}:
\begin{equation}\label{eqn:logn}
SFR(t)=\frac{1}{t\tau\sqrt{2\pi}}\cdot \exp{-\frac{(\ln t -t_0)^2}{2\tau^2}}
\end{equation}
where $t$ is the time in yr since the big bang, $t_0$ the  logarithmic delay time and $\tau$ the parameter that sets the initial SFR timescale.

The final model spectrum is obtained by summing the SSPs of different ages, each one weighted by a given stellar mass value calculated from the SFH. For the sake of simplicity, we did not include dust extinction. The age of the SSP goes from $10^4$ to $T_{max}$, where the latter is calculated based on the age of the Universe at the model's redshift, and assuming a galaxy formation redshift of 20. We choose a model redshift of 0.06, similar to that of the galaxies in the sample, which gives $T_{max}=12.25$ Gyr in the adopted cosmology. 

\begin{figure}
\centering
\includegraphics[scale=0.2]{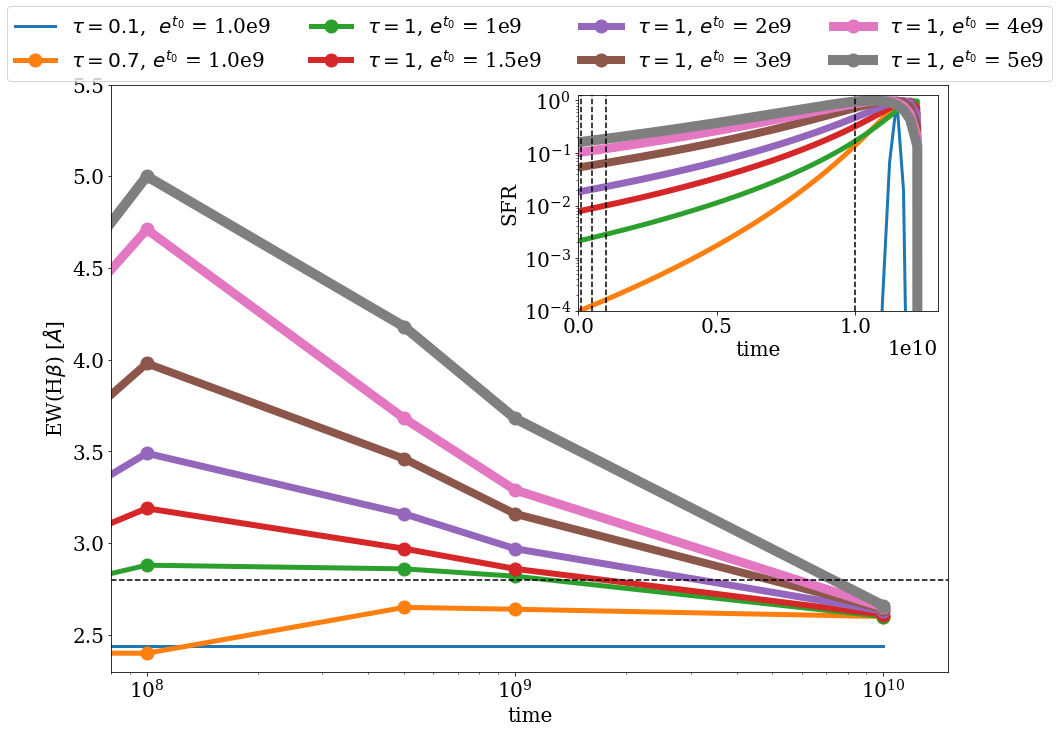}
\caption{EW(\Hb) as a function of truncation time measured on a set of model spectra, obtained from a combination of the free parameters of a log-normal prescription that allows us to simulate SFH of a range of galaxies. The corresponding SFHs are shown in the inset. SFHs have been  truncated at different epoch (see vertical dashed lines) to simulate a sudden quenching at different epochs (see text for details). The horizontal dotted  line shows the value adopted to separate $k+a/a+k$ from $k$ spectra.  \label{fig:models} }
\end{figure}

We explored model spectra obtained from a combination of the free parameters of the log-normal prescription that allows us to simulate SFH of galaxies across the whole Hubble sequence: from elliptical ($\tau=0.1$ and $t_0=20.7$, representing a short burst at the formation epoch), to actively star forming spiral (e.g. $\tau=1.0$ and $t_0=22.1$). Furthermore, for each of the SFHs, we have created a similar model but applying also a truncation in the SFR at epochs of $10^8$, $5\times 10^8$, and $10^9$ years ago, to simulate a sudden quenching at different epochs. Truncation at  $10^{10}$ yr ago replicates the spectrum of elliptical galaxies. { A schematic view of the different set of models along with the different parameters is presented in the label of Fig. \ref{fig:models}}

We then measured the EW(\Hb) on the model spectra and found that SFHs representative of elliptical galaxies produce spectra with EW(\Hb)$\sim 2.5$ \AA{}, while SFHs representative of actively star forming galaxies have, as expected, \Hb in emission. Spectra of star forming galaxies whose star formation has been truncated have instead \Hb in absorption whose magnitude depends on the epoch of the truncation. Fig. \ref{fig:models} shows that star forming galaxies { (represented by model SFHs with $\tau=1.0$)} whose star formation was truncated around $10^9$ yr ago or { even more recently} have always  EW(\Hb)$\geq 2.8$ \AA{} {(corresponding to the green point at t=$10^9$ yr in Fig. \ref{fig:models})}. This value can be adopted as a lower limit to identify galaxies in a post-starforming phase. Galaxies having a burst prior to quenching will have even larger EW(\Hb) values. 

We are now in the position of interpreting the integrated EW values shown in Table \ref{tab:EW}:   A3128\_B\_0248, and  A3158\_B\_0223 have  EW(\Hb) measured $<2.8$ \AA{}, while A1069\_B\_0103, A3158\_11\_91 and A500\_22\_184 have a value slightly above the threshold. The EW(\Hb) is therefore consistent with a scenario where the star formation of the former galaxies ended more than $\sim10^{9}$ yr ago,
that in the latter galaxies slightly earlier.

The rest of the sample is characterised by EW(\Hb) between 3.1 and 5.2\AA{}, consistent with populations quenched few $10^8$ yr ago and possibly having experienced a burst prior to the truncation.

\begin{figure}
\centering
\includegraphics[scale=0.26]{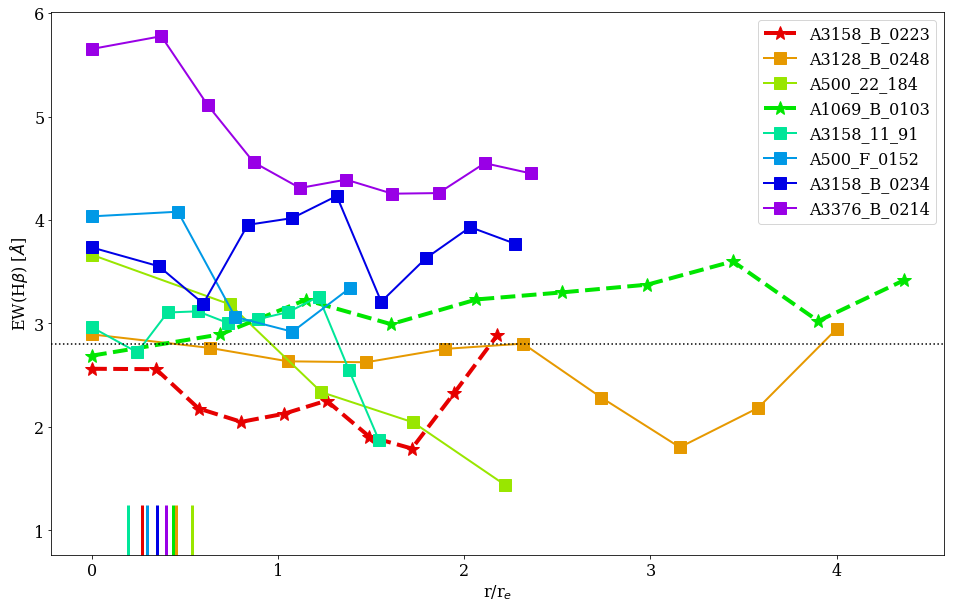}
\caption{Rest frame  EW(\Hb) gradients in units of r$_e$, for each galaxy of the sample. A small horizontal shift has been applied to the points for display purposes. Galaxies represented by stars have $k$ integrated spectra, galaxies represented by  squares  have $k+a/a+k$ integrated spectra. The thick vertical lines on the bottom indicate the size of 1 kpc in unit of $r_e$ for each galaxy.  The horizontal dotted  line shows the value adopted to separate $k+a/a+k$ from $k$ spectra. \label{fig:Hb_gradients} }
\end{figure}

For each galaxy, we can also characterise spatial trends in \Hb as a function of the galactocentric distance. To reduce the noise, we  compute EWs on stacked spectra. We consider annuli, defined as the regions in between elliptical apertures, which are chosen to match roughly the surface brightness intensity of the stellar emission at different levels. For all galaxies we consider 10 equally spaced annuli, but for the two smallest galaxies, for which we consider only five annuli. 

Figure \ref{fig:Hb_gradients} shows the EW as a function of the  galactocentric distance,  
normalised by $r_e$. 
Different trends can be seen and galaxy-by-galaxy variations are observed.
EW(\Hb) is lower than 2.8\AA{} at all distances for  A3128\_B\_0248, and  A3158\_B\_0223. These galaxies are therefore truly passive at all galactocentric distances. A1069\_B\_0103 and  A3158\_11\_91 have EW(\Hb) $<$3\AA{} in the galaxy center, then values are slightly higher in the outskirts, but never exceed EW(\Hb)$\sim 3.5$\AA{}. 
Overall trends are rather flat.  A500\_22\_184 is instead characterised by a very steep negative gradient. The EW(\Hb) is 3.8\AA{} in the galaxy center and then reaches values $<2$\AA{} in the outskirts. The galaxy could therefore have been passive for longer time in the outskirts and just recently become quickly passive in the core or could have had experienced a recent increase in the SF within the core.

The other three galaxies have  EW(\Hb) $>>2.8$\AA{} at all distances from the center. They also show higher EW(\Hb) values in the cores. A3376\_B\_0214 is  the galaxy  with the overall highest EW and also the highest EW in the center, with values ranging between 4.1 and 5.9\AA{}, followed by  A500\_F\_152 that also shows a strong EW(\Hb) gradient. In contrast,  A3158\_B\_0234 presents a flat trend. 

Figure \ref{fig:Hb_gradients} also shows the size of 1 kpc in unit of r$_e$, a region often inspected by other studies. The results will be discussed in \S\ref{sec:disc}.

To summarise the results so far, our sample includes two galaxies (A3128\_B\_0248,   A3158\_B\_0223) that have EW(\Hb)$<$2.8 \AA{} throughout the entire galaxy disks, indicating they have stopped forming stars at early epochs. These galaxies  are also the two reddest galaxies in the sample (see Fig.\ref{fig:cf_P17}) and are among the most massive galaxies. 
The sample also includes three galaxies (A500\_F\_0152, A3158\_B\_0234 and A3376\_B\_0214) whose EW(\Hb) measured on the integrated spectra is $>>$2.8 \AA{}, suggesting a recent truncation of their star formation ($<10^9$ yr ago). Two of these galaxies have strong negative EW(\Hb) gradients, suggesting the presence of a central burst before quenching. All these galaxies still show blue colors in Fig.\ref{fig:cf_P17}. The remaining three galaxies
have EW(\Hb)$\sim 3.1$\AA{}. A1069\_B\_0103  and A3158\_B\_0234 have flat EW(\Hb) gradients and green colors, while A500\_22\_184 have a steep negative gradient (EW(\Hb)$\sim 3.8$\AA{} in the core)  but redder colors. The last galaxy could therefore have been passive in the outskirts for a longer time, but have had an abrupt truncation of the star formation in the core recently. 

\begin{figure*}
\centering
\includegraphics[scale=0.35]{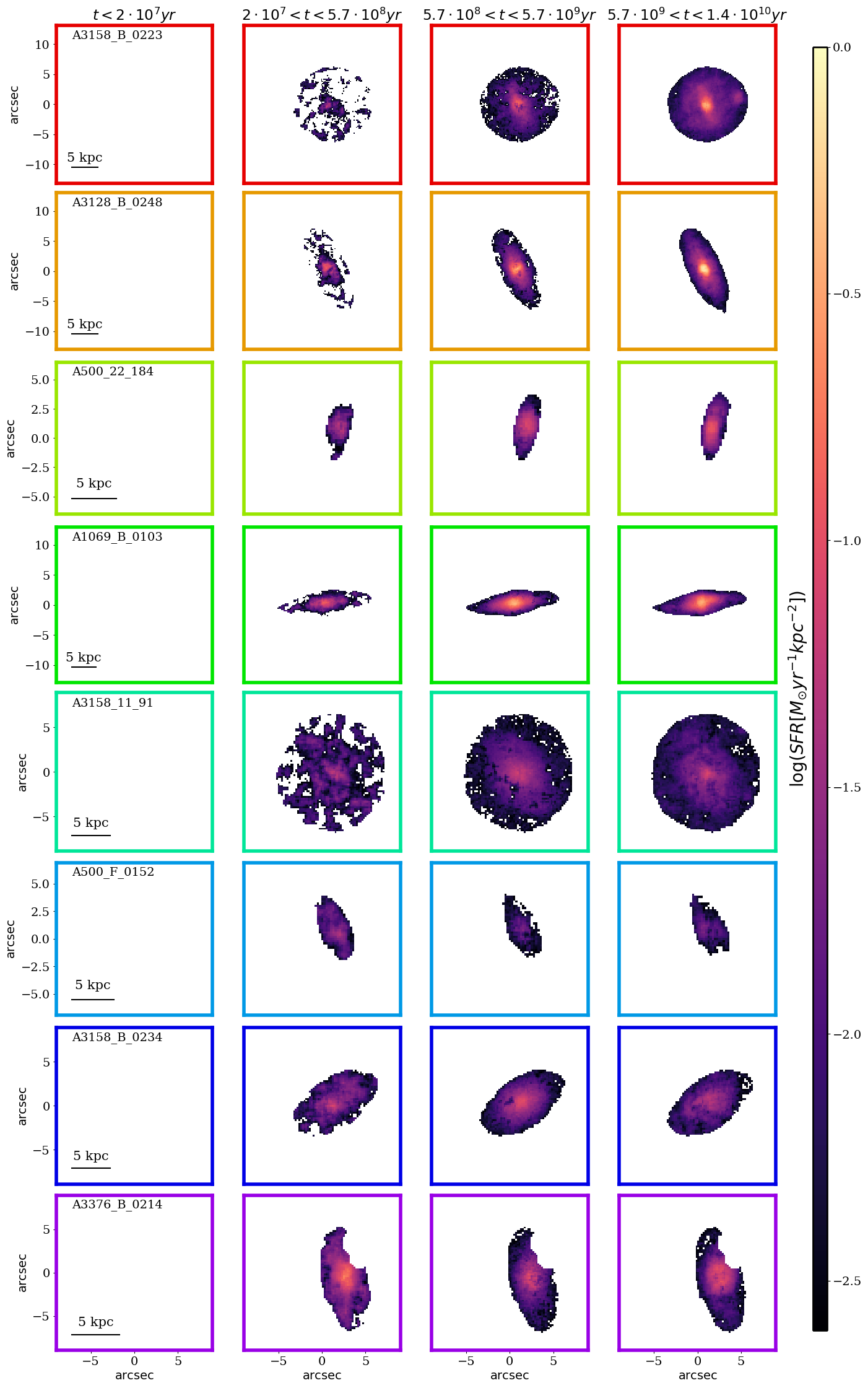}
\caption{Stellar maps of different ages, illustrating the average star formation rate per kpc$^2$ during the last $2\times 10^7$ yr (left), between $2\times 10^7$ yr and $5.7 \times 10^8$ yr (central left), $5.7 \times 10^8$ yr and $5.7 \times10^9$ yr (central right) and $> 5.7 \times 10^9$ yr ago (right), for each galaxy, separately. Galaxies are surrounded by squares colored following the scheme of Fig.\ref{fig:cf_P17}.{ Galaxies are rainbow-colored by increasing EW(\Hb), as measured on the integrated spectra (see Tab.\ref{tab:EW})}. \label{fig:PSB_SFH_4bins}}
\end{figure*}

\section{Spatially resolved properties}\label{sec:results2}

\subsection{Star formation Histories}

A further characterisation of the quenching process these galaxies underwent can be obtained inspecting their spatially resolved SFHs{ , obtained using \sinopsis}.

\begin{figure}
\centering
\includegraphics[scale=0.45]{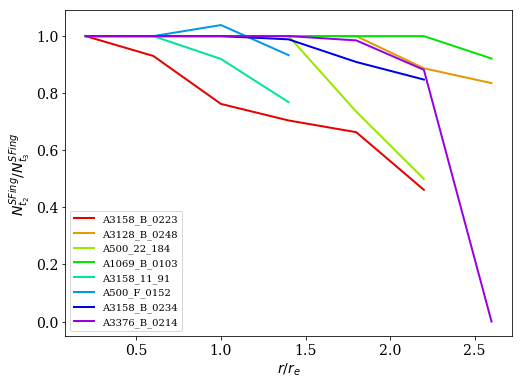}
\caption{{ Ratio of the number of spaxels that were star forming in the second age bin ($t_2=2.7 \times 10^7$yr - $5.7 \times 10^8$yr ago) to the number of spaxels that were star forming in the third age bin ($t_3=5.7 \times 10^8$yr - $5.7 \times 10^9$yr ago), as a function of the galactocentric distance.  \label{fig:PSB_SFH_ratios}}}
\end{figure}

Figure \ref{fig:PSB_SFH_4bins}  shows  the variation of the SFR maps with cosmic time in four age bins. Given that no emission lines are detected, none of the galaxies have star formation in the youngest age bin (last $2\times 10^7$ yr), consistent with the fact that they are all truly passive today. 
All the galaxies have stopped forming stars at least 20 Myr ago { and were highly star forming in the oldest age bin. In contrast, they present different characteristics in the two intermediate bins.} A3376\_B\_0214 is the only galaxy showing a clear sign of a central SF enhancement between $2.7 \times 10^7$yr and $5.7 \times 10^8$yr ago. The three galaxies with the highest EW(\Hb) (A500\_F\_0152, A3158\_B\_0234 and A3376\_B\_0214) show 
{ only mild} signs of { star formation truncation in the galaxy outskirts} in the second age bin (between $2.7 \times 10^7$yr and $5.7 \times 10^8$yr ago) { compared to the third age bin (between $5.7 \times 10^8$yr and $5.7 \times 10^9$yr ago), indicating that the bulk of the quenching process occurred more recently than $2.7 \times 10^7$yr ago. 
 In contrast, the other galaxies show clearly that the quenching process started at earlier epochs and proceeded from the external regions towards the galaxy cores (outside-in quenching). The outside-in quenching is better observed in Fig.\ref{fig:PSB_SFH_ratios}, which shows the ratio of the number of spaxels that were star forming in the second age bin to the number of spaxels that were star forming in the third age bin, measured in annuli at different galactocentric distances. These are the two age bins that better capture the quenching: at older epochs all spaxels were star forming, in the youngest age bin the galaxies are completely quenched. The figure shows that in the galaxy cores all spaxels are star forming in both age bins, while as we proceed from the galaxy cores toward the outskirts the fraction of star forming spaxels decreases, indicating that a larger portion of the galaxies is already quenched.  A3158\_B\_0223 is the galaxy with the strongest evidence of outside-in quenching, while the trends for A328\_B\_0248 and A1069\_B\_0103 have to be taken carefully because the former galaxy is very small and so there are not many spaxels in each galactocentric distance bin and the latter has a high inclination and annuli enclose portion of the galaxies that might be physically different.}

\begin{figure*}
\centering
\includegraphics[scale=0.5]{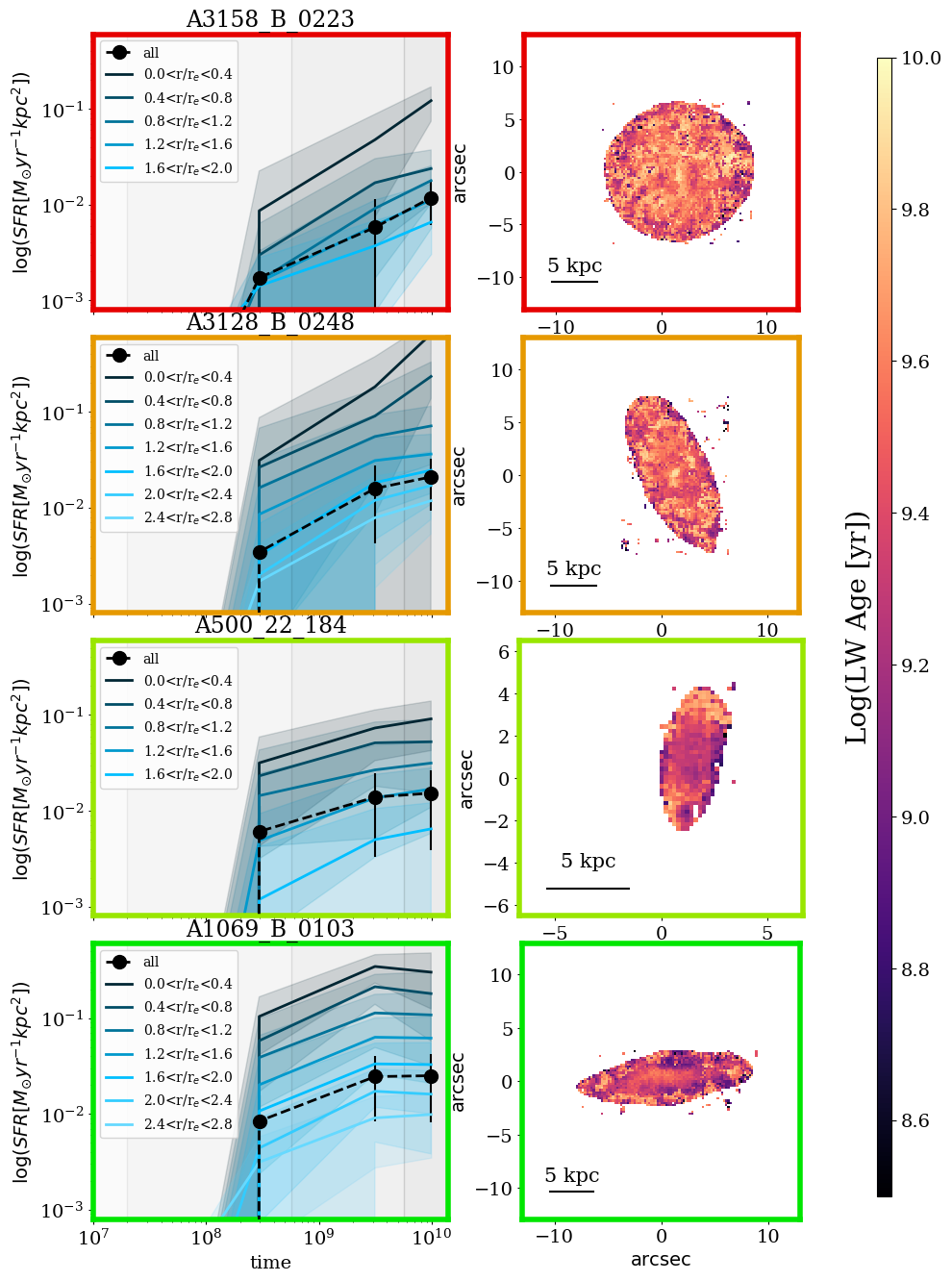}
\caption{{ Left}: SFHs  in equally spaced different regions of 4 galaxies of the sample. { The number of regions depends on the extension of the galaxy and is indicated in the label.} Black dots and dashed lines represent the SFH of the galaxy across the entire galaxy disk. Shaded areas and black lines show the 1$\sigma$ dispersion of the curves. In the background, grey areas identify the width of the bins. Right: Maps of luminosity weighted ages. Galaxies are surrounded by squares colored following the scheme of Fig.\ref{fig:cf_P17}. { Galaxies are rainbow-colored by increasing EW(\Hb), as measured on the integrated spectra (see Tab.\ref{tab:EW})}.
\label{fig:PSB_SFH_4bins_LWA} }
\end{figure*}

\begin{figure*}
\centering
\includegraphics[scale=0.5]{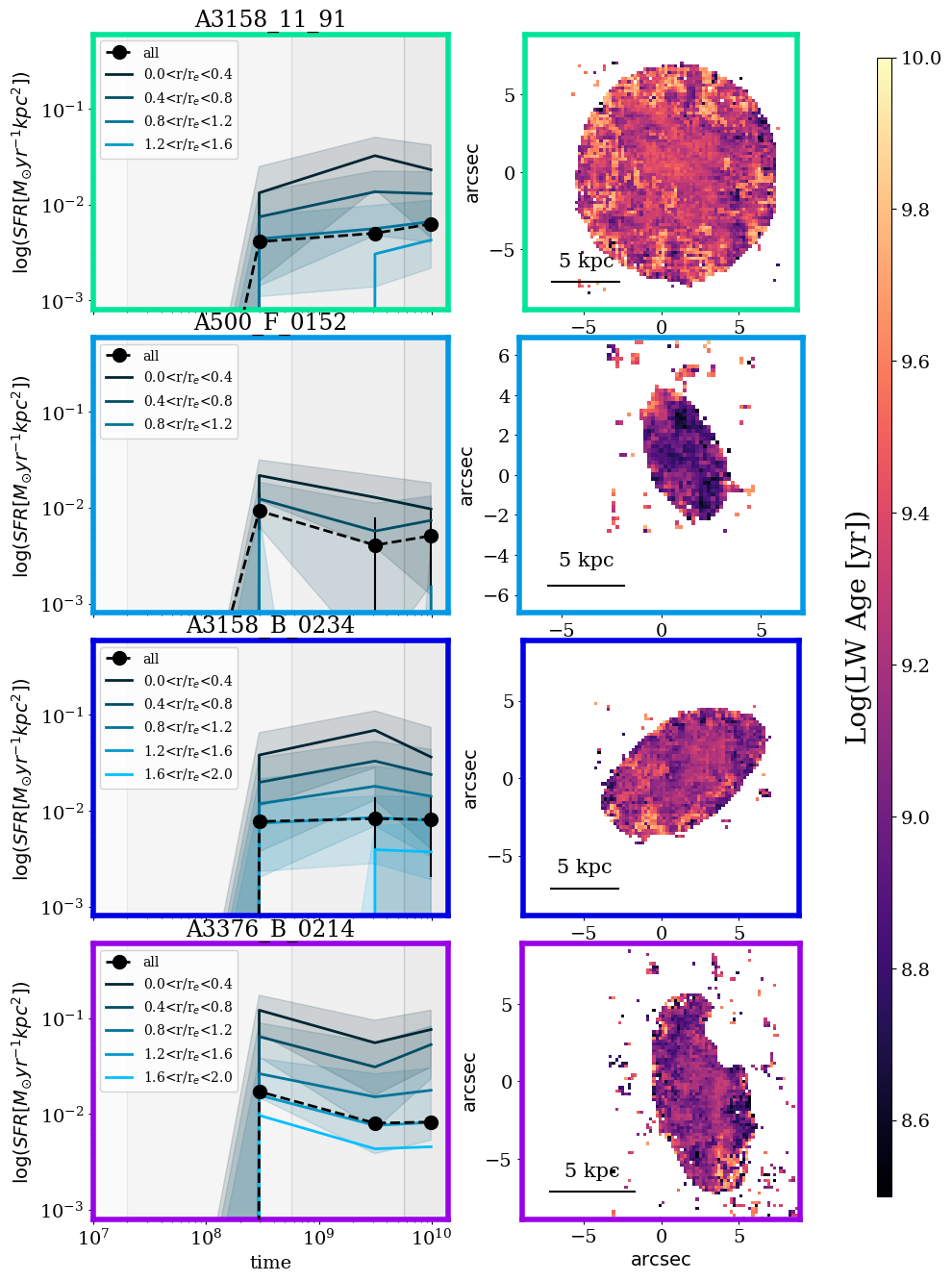}
\caption{Same as Fig.\ref{fig:PSB_SFH_4bins_LWA} for the other four galaxies of the sample. \label{fig:PSB_SFH_4bins_LWA2}}
\end{figure*}

The left panels of Fig. \ref{fig:PSB_SFH_4bins_LWA} and \ref{fig:PSB_SFH_4bins_LWA2}, show more in detail the SFHs, both integrated over the whole galaxy disks
and in different regions of the galaxies. Trends are shown with errors, which are computed as  standard deviation on the plotted median values. Uncertainties on the SFR measured on each spaxel are negligible.\footnote{{ Following the approach explained in \citet{Fritz2007}, for each spaxel we have run \sinopsis 11 times for each chosen value of the SSP metallicity: the final results have been used to get lower and upper limits for the SFR in the different age bin. When these are binned not only temporally but also spatially to get radial profiles, we have found that the dominant source of uncertainty is the spaxel-to-spaxel variation, that we have reported in the plots. We explained this in the text.}}

Considering integrated values, we observe that even though all galaxies are quenched today, 
 they reached the passive state through different paths.

The two $a+k/k+a$ galaxies with strong EW gradients (A500\_F\_0152 and A3376\_B\_0214) show an increase of the SFR in the second age bin before the quenching- suggesting again an enhancement  before the truncation of the star formation,  while A3158\_B\_0234 presents a rather flat trend. In contrast,  all the other galaxies show a steady decline of the SFR from the oldest towards the youngest ages. 

Moving the attention to portions of galaxies at different galactocentric distances, we observe that, at any given epoch the median SFR is always the highest in the core and then it decreases towards the outskirts, in a monotonic manner. { This behaviour is partially due to the fact that the stellar mass is concentrated to the center: if we inspect the sSFHs (plot not shown) we obtain a smaller variation with distance. The remaining  scatter indicates that t}he galaxy outskirts have been always less effective in forming new stars. 

Secondly, we observe that the SFHs at a given distance do not follow exactly the integrated value indicating that different portions of the galaxy assembled their mass differently. In particular, the outskirts of A3158\_11\_91, A3128\_B\_0248 and A3158\_B\_0223 became passive between $5.7\times10^8$ and $5.7\times10^9$ yr ago.


\begin{figure}
\centering
\includegraphics[scale=0.25]{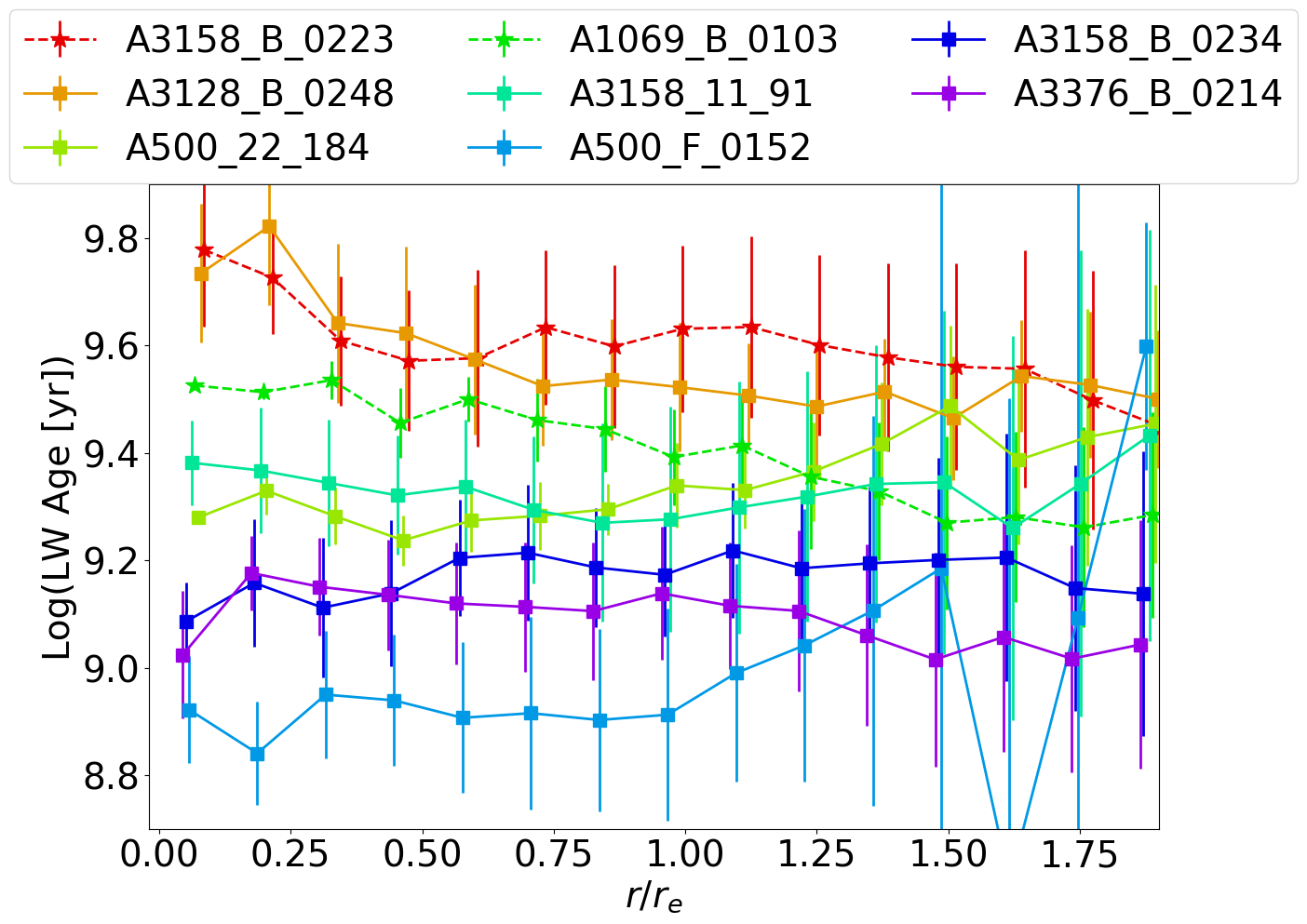}
\caption{Luminosity weighted age gradients in unit of r$_e$, for each galaxy of the sample. Errors represent the standard deviation (1$\sigma$). A small horizontal shift has been applied to the points for display purposes. Galaxies represented by stars have $k$ integrated spectra, galaxies represented by  squares  have $k+a/a+k$ integrated spectra. 
\label{fig:psb_LWA_gr} }
\end{figure}

The right panels of Fig. \ref{fig:PSB_SFH_4bins_LWA} and \ref{fig:PSB_SFH_4bins_LWA2} show the 
luminosity-weighted age maps, that provide  an estimate of the average age of the stars weighted by the light, therefore giving us an indication of when the last episode of star formation occurred.
An age spread among galaxies is detected, with galaxies showing $a+k/k+a$ spectra having systematically younger ages (between $8.8<\log (age \, [yr])<9.2$) than the other galaxies (between $9.2<\log (age \, [yr])<9.8$). No particularly strong radial variation with distance are observed. 
Figure \ref{fig:psb_LWA_gr} shows in detail the variation of the luminosity-weighted ages with distance, in units of $r_e$, similarly to what shown in Fig.\ref{fig:Hb_gradients} for the EW(\Hb). { The Figure shows a mix of flat, positive, and negative gradients, which might suggest a mix of outside-in, inside-out, and simultaneous quenching. In particular,}
A3376\_B\_0214 - the galaxy with the most outstanding EW(\Hb) negative gradient - presents a dip in the LWA gradient in the core, indicative of a younger age, and then a rather flat gradient. 
Similarly, also the other galaxies with $a+k/k+a$ spectra and A500\_22\_184 show positive gradients, suggesting the presence of younger ages in the cores. 
{ A500\_F\_0152  shows a significant younger age at $r/r_e<$1  than $r/r_e>$1. This is consistent with what found in Fig.\ref{fig:Hb_gradients}, where this galaxy showed a very steep EW(\Hb) gradient.}

The other galaxies present overall flat or even negative gradients. 
These results are supported by the Pearson r-correlation test { (see Tab. \ref{tab:Pear})}. 
{ Results of Fig.\ref{fig:Hb_gradients} and Fig. \ref{fig:psb_LWA_gr} are therefore consistent and show how the EW(\Hb) is a good age tracer.}

\begin{table}
\caption{Pearson r-correlation test results performed on the luminosity-weighted age gradients.  \label{tab:Pear}}
\centering
\begin{small}
\begin{tabular}{l|c | c }
  \multicolumn{1}{c|}{id} &
  \multicolumn{1}{c|}{coefficient} &
  \multicolumn{1}{c}{p-value}\\
\hline
  A500\_22\_184 & 0.85 & 6e-05 \\
  A500\_F\_0152 & 0.5 & 0.07\\
  A3158\_B\_0234 & 0.39 & 0.1\\
  A3376\_B\_0214 & -0.6 & 0.01 \\
  A3376\_B\_0214* & 0.2 & 0.6 \\
  A3158\_11\_91 & -0.006 & 1\\
  A3128\_B\_0248 & -0.7 & 0.0006\\
  A1069\_B\_0103 & -0.96 & 6e-09\\
  A3158\_B\_0223 & -0.80 & 0.0003 \\
\end{tabular}
\tablenotetext{*}{values computed considering data within 1 $R_e$, to better catch the dip in the galaxy center.}

\end{small}
\end{table}

\section{Discussion}\label{sec:disc}

In this paper we have characterised the spatially resolved properties of six galaxies showing $k+a/a+k$ fiber (central $\sim$ 2 kpc) spectra and two having $k$ fiber spectra. The goal is to understand their pathway to the quenching state, what are the mechanisms that affected their star formation activity and on what timescales. 

The first important result is that galaxies are entirely quenched, and no signs of emission lines are detected throughout the galaxy disks. { As these galaxies were selected on the basis of fiber spectra, it could have been that only the central regions of the galaxies were characterized by $ k+a/a+k$ features, while star forming regions were still present in the outskirts.} 

We investigated the integrated properties and the spatial distribution of the stellar populations to obtain information about the mechanism responsible of the galaxy features. 

The analysis of the surface brightness profiles  has shown that most of the galaxies are characterised by the presence of { either } bars{ , lens or} truncated Type II disks. Head et al. (2015) have found that the bar fraction is considerably higher in galaxies hosting a  Type II disk than in galaxies whose disc remains unbroken. This implies that either the truncation  mechanism  induces bar growth, or that bars stabilise discs during truncation, such that the detection of a disc break for bright galaxies is more likely if a bar is present. The significant increase in bar size for more luminous Type II galaxies may therefore suggest a period of enhanced star formation in the bar due to gas inflows, visible as a burst. Even regardless of the presence of a bar, truncated discs alone are also  indicative of an episode of star formation which lasted longer in the galaxy center than in the outskirts: the steeper profile of the inner stellar disk is likely due to the inside-out grow mode of the disks (Elmegreen \& Parravano 1994, Martinez-Serrano et al. 2009).

Central bursts and rapid cut off of the 
 star formation { can} also leave a young stellar population in a centrally concentrated cusp \citep[with scales of $\sim$1 kpc,][]{Bekki2005, Hopkins2009} and an old stellar population distributed like a normal early-type galaxy \citep{Hopkins2009, Snyder2011}. 

{ However, from the observational point of view, results regarding the spatial distribution of PSBs and the presence of stellar population gradients are still controversial. One one side, 
\cite{Norton2001,  Yagi2006, Chilingarian2009} found evidence the young stellar population is not confined to the galaxy core, but extends over $\sim$2-3 kpc; likewise \cite{Pracy2009} were unable to detect any Balmer line gradients or central concentration in the young population and \cite{Swinbank2012} claimed that the characteristic PSB signature is a property of the galaxy as a whole and not due to a heterogeneous mixture of populations.

On the other hand, \citep{Pracy2005, Snyder2011} did observed a Balmer line absorption enhancement and gradient in the central regions, interpreted as proxy for the existence of a young component in PSB galaxies. 
\cite{Pracy2013} have analyzed four low luminosity PSBs in different environments, including a cluster member and found that all  four galaxies do have centrally concentrated gradients in the young stellar population contained within the central $\sim$1 kpc.

In Fig. \ref{fig:Hb_gradients} we have shown the \Hb gradients for our sample and highlighted the 1 kpc size for each galaxy separately. In agreement with the results presented by \cite{Pracy2013}, the galaxies classified as $k+a/a+k$ show hints of centrally concentrated Balmer line gradients in the central 1 kpc, even though we are hitting the regime where results are dominated by the Point Spread Function. We note, however, that differently from the other studies, we could inspect gradient out to large galactocentric distances.

A reason for the mixing results is that different studies are based on different techniques (long slit spectroscopy vs IFU data), samples are very small (typically of the order of 10 galaxies at most), heterogeneous in terms of redshift range, environment, galactocentric distance probed, definition of PSB galaxies and have been hampered by physical scale resolution constraints.  Studies based on larger and more homogeneous samples  and with higher spatial resolution would be necessary to firmly determine the gradients. }

In the PSB galaxies this young component is expected to be observable as a Balmer line absorption enhancement and gradient in the central region \citep{Pracy2005, Snyder2011}. 
\cite{Pracy2013} have analyzed four low luminosity PSBs in different environments, including a cluster member and found that all  four galaxies do have centrally concentrated gradients in the young stellar population contained within the central $\sim$1 kpc.
In Fig. \ref{fig:Hb_gradients} we have shown the \Hb gradients for our sample and highlighted the 1 kpc size for each galaxy separately. In agreement with the results presented by \cite{Pracy2013}, the galaxies classified as $k+a/a+k$ show hints of centrally concentrated Balmer line gradients in the central 1 kpc, even though we are hitting the regime where results are dominated by the Point Spread Function.

\subsection{The formation scenario}
As mentioned in Sec.1, two main scenarios have been proposed to explain the formation of PSB galaxies. In the field, galaxy interaction and mergers are most likely the main mechanisms involved, while in clusters ram pressure stripping has been advocated. In the following, we will discuss the main observables that argue against or in favour to the two scenarios for our sample galaxies.

\subsubsection{The cluster environment}

The eight galaxies belong to five different clusters of the OMEGAWINGS survey. The main properties of the clusters are summarised in Table. \ref{tab:clus}. Clusters are characterised by different values of velocity dispersion ($550<\sigma [km/s]<950$), suggesting that galaxies might feel cluster specific processes with different strengths. A3376, A3158 and A3128 also belong to the Shapley superclusters, while A1069 and A500 are single, not merging systems. 

\begin{table}
\begin{center} 
\caption{Properties of the clusters hosting the PSBs: Cluster name, coordinates, redshifts, velocity dispersions and virial radius. Values are from \citet{Moretti2017, Biviano2017}.
\label{tab:clus}}
\begin{tabular}{l|r|r|r|r|r}
  \multicolumn{1}{c|}{Cluster} &
  \multicolumn{1}{c|}{R.A.} &
  \multicolumn{1}{c|}{DEC} &
  \multicolumn{1}{c|}{z$_{cl}$} &
  \multicolumn{1}{c|}{$\sigma_{cl}$} &
  \multicolumn{1}{c}{R$_{200}$}\\
  \multicolumn{1}{c|}{} &
  \multicolumn{1}{c|}{(J2000)} &
  \multicolumn{1}{c|}{(J2000)} &
  \multicolumn{1}{c|}{} &
  \multicolumn{1}{c|}{(km s$^{-1}$)} &
  \multicolumn{1}{c}{(Mpc)}\\
\hline
  A3376 & 90.1712 & -40.0444 & 0.04652 & 
  756$^{+39}_{-37}$ & 1.7$\pm$0.2\\
  A3158 & 55.71487 & -53.62543 & 0.05947 & 
  948$^{+48}_{-46}$ & 1.9$\pm$0.1\\
  A3128 & 52.4639 & -52.5806 & 0.06033 & 
  793$^{+40}_{-38}$ & 1.6$\pm$0.2\\
  A1069 & 159.9308 & -8.6867 & 0.06528 & 
  542$^{+36}_{-38}$ & 1.2$\pm$0.2\\
  A500 & 69.7187 & -22.11 & 0.06802 & 
  660$^{+33}_{-34}$ & 1.8$\pm$0.2\\
\end{tabular}
\end{center}
\end{table}

\begin{figure*}
\centering
\includegraphics[scale=0.29]{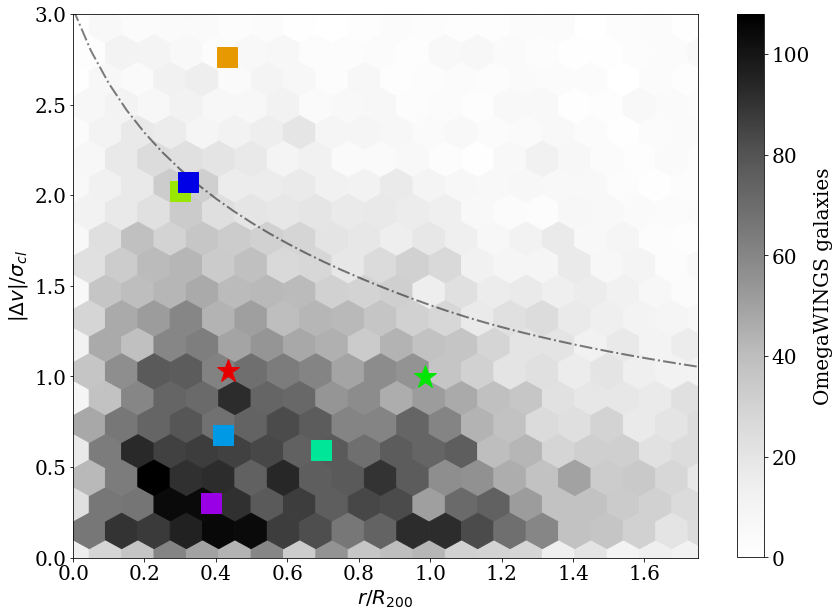}
\includegraphics[scale=0.29]{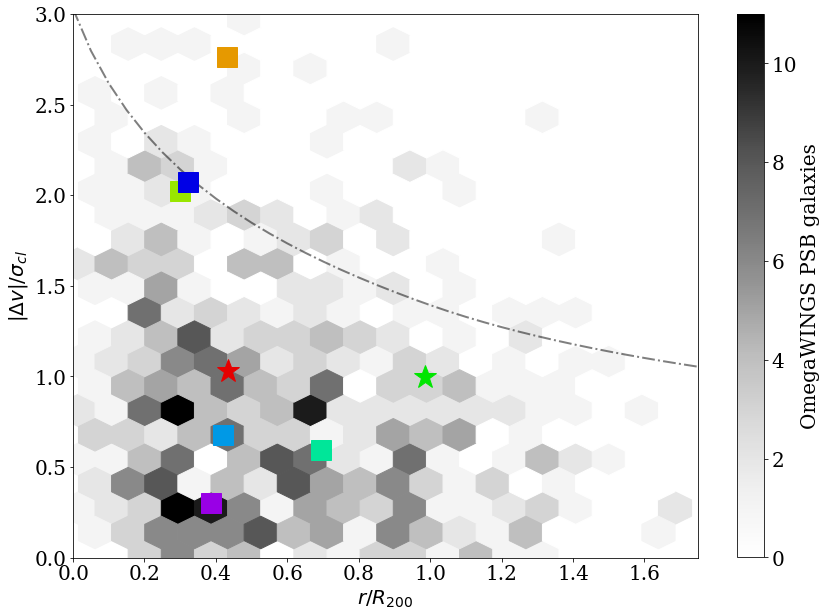}
\caption{Location in projected position versus velocity phase-space of  the PSB galaxies analysed in the paper. Galaxies represented by stars have $k$ integrated spectra, galaxies represented by  squares  have $k+a/a+k$ integrated spectra.  The background shows the distribution of all OMEGAWINGS clusters with spectroscopic completeness $>$50 per cent stacked together (left) and the distribution of  OMEGAWINGS PSBs from \citet{Paccagnella2017} stacked together (right, green color bar). The grey curve corresponds to the 3D (un-projected) escape velocity in a NFW halo with concentration c = 6 for reference. 
\label{fig:psb} }
\end{figure*}

Figure \ref{fig:psb} presents the galaxy locations in the so-called projected phase-space diagram. This plot compares the projected distance from the cluster center and the line-of-sight velocity relative to the cluster velocity, normalized by the cluster size and velocity dispersion, respectively. Generally,  galaxies at different positions in their orbit and with different times since infall occupy different regions in projected phase-space diagrams \citep{Oman2013}, which  could be associated with different amounts of tidal mass-loss \citep{Smith2015} and/or different strength of ram pressure stripping (\citetalias{Jaffe2018}, \citealt{Jaffe2015}).

According to cosmological simulations \citep{Haines2015, Rhee2017}, galaxies that have been in a cluster for a very long time have lower velocities and clustercentric radii because they have had enough time to sink into the potential well of the cluster. On the other hand, infalling galaxies approach the cluster core with high relative velocities at all clustercentric distances. Observations of galaxies in clusters confirm this scenario \citep{Jaffe2015, Yoon2017}.

All our galaxies are located between 0.3 and 1 $R_{200}$, therefore avoiding the cluster cores, but they are characterised by quite large a range of
relative velocities (0.2-2.8 $|\Delta(v)|/\sigma$), indicating they are moving towards or away from  their cluster centers with different speeds. 

A3128\_B\_0248 is the galaxy with the highest relative velocity ($|\Delta(v)|/\sigma$=2.8), but it shows passive features. It might therefore be either a pre-processed galaxy or an ancient infaller, rather than a recent infaller. \cite{Rhee2017} have indeed shown that $\sim$ 40\% of galaxies in this area are ancient contaminants to the recent infaller population. A500\_22\_184 and A3158\_B\_0234 are very close to the edge of the typical region delimited by the escape velocity in relaxed systems, and given their spectral properties they are consistent with being recent infallers that suffer a truncation of the star formation as a consequence of their first encounter with the harsh cluster environment. 

The other galaxies are located inside the typical trumpet-shaped region of the relaxed systems, in an area where both recent and ancient infallers could be actually found \citep{Rhee2017}. 

We remind the reader that galaxies in A3376, A3158 and A3128 are part of the Shapley supercluster, therefore their position in the phase space need to be treated with caution as cluster mergers can displace galaxy positions. 

In Fig. \ref{fig:psb} galaxies are overplotted on the projected phase space diagram of all OMEGAWINGS galaxies (left) and on all the OMEGAWINGS PSB galaxies from \cite{Paccagnella2017} (right). 
2d k-s tests state that the galaxies analysed in this paper have a different distribution from the total population of cluster members, while it is indistinguishable from the entire PSB population.
The position of most of them 
in the phase-space diagram is 
consistent with a recently infalled population that have been into clusters for at least one pericenter passage (i.e. $\geq$1Gyr). 
These galaxies could therefore be descendants of  galaxies that quickly lost the gas when they were on first infall due to ram pressure stripping (i.e. jellyfish galaxies).

While the galaxies presented here have no star formation throughout the disk, GASP has been very successful in identifying also PSB regions in galaxies currently undergoing ram pressure stripping (\citetalias{Gullieuszik2017,Poggianti2019b}),  highlighting how rapid events exhausting the gas due to ram pressure do produce PSB spectra.

\begin{figure}
\centering
\includegraphics[scale=0.35]{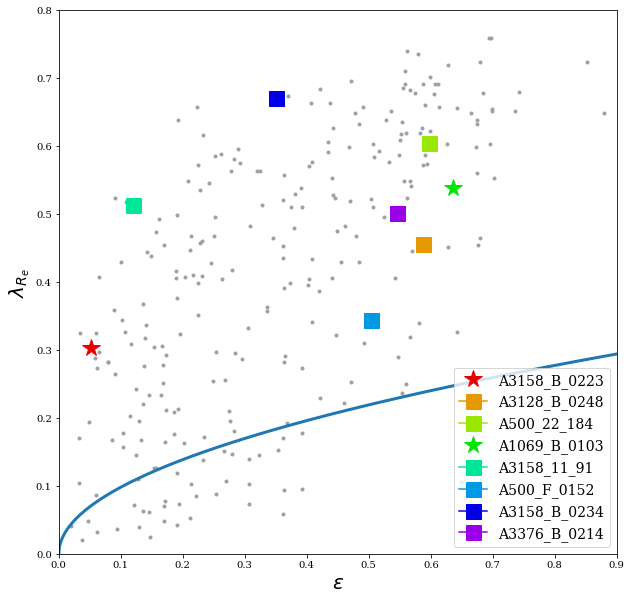}
\caption{$\lambda_{R_e}$ versus the ellipticity $\epsilon$ for the sample galaxies (red dots and squares) compared with
ATLAS3D sample (grey dots, from \citealt{Emsellem2011}). Slow and fast rotators are split by the blue line. 
\label{fig:lambda}}
\end{figure}

\subsubsection{Ruling out galaxy mergers}

The projected stellar angular momentum per unit mass, $\lambda_R$ parameter (Emsellem et al. 2007, 2011), is commonly used to quantify the kinematic state of early-type galaxies. 
Emsellem et al. (2007) showed that  galaxies can be separated into two distinct kinematic classes depending on the value of $\lambda_R$: the fast and slow rotators. Most of the early-type galaxies are classified as fast rotators \citep{Emsellem2011}, as also $k+a/a+k$ galaxies (Pracy et al. 2009, 2012; Swinbank et al. 2012). The dearth of slow rotators in the E+A population has been used to argue against the need for major galaxy mergers in their production \citep{Pracy2009, Pracy2012}, since the probability of a rotating remnant increases as the mass ratio of the progenitors involved in the merger increases  \citep{Bournaud2008} and major mergers should result in an increased fraction of slow rotators \citep[see also][]{Graham2019}.

Fig.\ref{fig:lambda} shows the $\lambda_R$ parameter vs. ellipticity for the sample. As reference, the datapoints of the 260 early-type galaxies in the nearby Universe  of ATLAS$^{3D}$ from \cite{Emsellem2011} are also shown. All the galaxies are clearly fast rotators. This piece of evidence, together with the regular morphologies (Fig. \ref{fig:color_images}) and stellar kinematics (Fig. \ref{fig:vel}) and the characterisation of the environment, argue against the merger scenario.

In addition, we have also more carefully investigated the local environment of the 8 galaxies, to look for signs for possible merger features. In particular we checked whether 1)  they have companions 2) they are in substructures 3) they are in high density regions. 

We searched in the entire OMEGAWINGS sample candidate companions with measured redshift and measured galaxy properties. Only in two cases we have found suitable candidates:  A500\_22\_184, the least massive system in our sample, has a massive ($M_\ast=7\times 10^{10} M_\sun $, a factor of 200$\times$ larger) passive galaxy at $\sim 80$ kpc and with a velocity difference of $\sim 135$ \kms. 
A3158\_B\_0223, the most massive object in our sample,  has a close companion at 40 kpc. The velocity difference between the two is  $\sim 135$ \kms. The companion is also passive and has a similar stellar mass ($M_\ast =  2.6 \times 10^{10} M_\sun $). The same galaxy is also the only galaxy which is most likely part of substructure in its cluster. Substructures have been defined running the DS+ method of \citet{Biviano2017} on the cluster members.

The projected local galaxy densities (LD) were derived calculating the number of galaxies per  $\rm Mpc^{2}$ \citep[see][for details]{Vulcani2012}, using the 10th nearest neighbours with $M_V  \leq -19.5$.  The galaxy local densities range between $0.97<\log (LD [Mpc^{-3}])< 1.5$. As reference, the median  project local density cluster members above $3.5 \times 10^9$ \ma/\ms (the OMEGAWINGS mass completeness limit) is 1.2$\pm$0.2. Therefore these galaxies are found in averagely dense regions and not in the most extreme ones. 

These pieces of evidence also argue against the merger scenario.

\section{Summary and Conclusions}

GASP is a project developed to understand gas accretion and removal processes in the different environments, with the aim to 
determine
the role of the environment in shaping galaxy properties. 

In this paper we have presented the characterisation of eight galaxies, included in the GASP sample as representative of the final stage of the evolution of galaxies in clusters. 

The combination of integrated and spatially resolved properties of the galaxies have allowed us to divide galaxies into three groups.

A3158\_B\_0234,  A3376\_B\_0214 and  A500\_F\_0152 are $k+a/a+k$ galaxies: they are located in the blue region of the color magnitude diagram,  have EW(\Hb)$>$2.8\AA{} both when considering the integrated spectrum and when considering stacked spectra representative of different portions of the galaxies. Their LWA are $<10^{9.2}$ yr at all galactocentric distances, their SFH show signs of a rapid truncation. A500\_F\_0152 and  A3376\_B\_0214 also show hint of a central SF enhancement just before quenching, 
A3158\_B\_0234 does not. 
The galaxies are located in almost the same region of the projected phase space diagram: relatively low velocities  ($<1$$|\Delta(v)|/\sigma$), intermediate clustercentric distances, where relatively recent infallers can be found \citep{Rhee2017}.

A3158\_B\_0223 and  A3128\_B\_0248 are very red, have $k$ spectra at all distances from the center and declining SFH since $10^{10}$ yr ago. They show the oldest luminosity weighted ages ($>10^{9.6}$yr).
All their properties indicate they have been fully quenched more than 1 Gyr ago. Given their stellar mass, these galaxies (as also A1069\_B\_0103) might have been quenched while still in the field, { or in pre-processing groups}.  

The remaining three galaxies have properties intermediate between the aforementioned groups, with intermediate colors, intermediate LWA,  intermediate EW (\Hb) and less steeply declining SFH. Of these, A500\_22\_184 stands out, because even though it has an overall quite low global EW(\Hb), it has a very steep \Hb gradient, suggesting the presence of a central burst. The galaxy can not be conclusively classified as $a+k/k+a$ because the truncation of the star formation in the outer disk regions occurred more than 10$^9$ yr ago, but the central regions still show $a+k/k+a$ signatures. 

The three groups are probably not distinct classes, but represent an evolutionary sequence cluster galaxies can undergo: after entering the cluster environment, galaxies abruptly quench as an effect of gas removal due to ram pressure stripping. At first, these objects maintain their morphology and blue colors, but their spectral properties immediately change: emission lines disappear and deep absorption lines emerge. If a burst of star formation occurred prior to quenching, absorption lines are even deeper and EW gradients are visible. As time goes by, 
galaxy populations age and become older, EW of absorption lines decrease and as a consequence galaxy colors become redder. After $>$10$^{9}$ yr,
EW of absorption lines are minimal and galaxies are red. All these changes occur before the galaxy morphology changes, suggesting that morphological transformations occur on even longer time scales. Nonetheless, morphological transformations can occur, as witnessed by the presence of S0s in our sample. These observations support a scenario in which at least some S0s can form via ram pressure stripping (and perhaps additional mechanisms) in clusters.

{  Most of the galaxies have been quenched outside-in, i.e. the outskirts reached undetectable SFRs before the inner regions, even though this feature is more visible for the $k$ galaxies (see Fig. 9).}

The outside-in quenching, the lack of signs of interactions and the high $\lambda_R$ measured, together with the fact that the galaxies are found in dense clusters, point to a scenario according to which ram pressure stripping has removed the gas. 

{ It is interesting to observe that actually ram pressure stripping can produce a variety of spatial trends in SFH during quenching.}

While in this paper we have focused only on the optical properties of the galaxies, to get a complete understanding of the baryonic cycle in PSB galaxies a multi-tracer study covering atomic, and molecular, in addition to ionised gas would be very important, similarly to what performed in \cite{Klitsch2017} for field galaxies, to quantify the state and distribution of all the gas components and predict whether these galaxies still have a substantial reservoir of atomic and/or diffuse molecular gas,
that could eventually form stars again.

\acknowledgments

We thank the referee for their comments that helped to improve the manuscript. Based on observations collected at the European Organisation for Astronomical Research in the Southern Hemisphere under ESO programme 196.B-0578. This project has received funding from the European Reseach Council (ERC) under the Horizon 2020 research and innovation programme (grant agreement N. 833824). We acknowledge funding from the INAF PRIN-SKA 2017 program 1.05.01.88.04 (PI Hunt). We acknowledge financial contribution from the contract ASI-INAF n.2017-14-H.0, from the grant PRIN MIUR 2017 n.20173ML3WW\_001 (PI Cimatti) and from the INAF main-stream funding programme (PI Vulcani).  Y.~J. acknowledges support from CONICYT PAI (Concurso Nacional de Inserci\'on en la Academia 2017) No. 79170132 and FONDECYT Iniciaci\'on 2018 No. 11180558.

%

\vspace{5mm}
\facilities{VLT(MUSE)}


\software{}



\appendix
\section{Surface Brightness profiles}\label{app:SBP}
As explained in detail in Franchetto et al. (in prep.), we perform an isophotal analysis on the I-band images using the {\sc ellipse} task in {\sc iraf} \citep{Jedrzejewski1987} to extract the Surface Brightness profiles (SBPs) of the galaxies. {\sc ellipse} fits on the image a series of elliptical isophotes such to minimize the deviations from the real shape of the galaxy isophotes. Then it returns the mean intensity along the ellipse, semi-major axis, position angle and ellipticity ($\varepsilon$) for each one. We mask out foreground stars, nearby and background galaxies, bad pixels, and bright spots before fitting the isophotes. 

In order to investigate the photometric structures that contribute to the galaxy light distribution, we carry out a parametric monodimensional decomposition of the observed SBPs. 

The components and the corresponding photometric laws considered to perform the best fit decomposition are the following:
\begin{equation}
    \text{Bulge:}\qquad I(r)=I_{\rm e}\,10^{-b_n\,[(r/r_{\rm e})^{1/n}-1]}\qquad \text{\citep{sersic1968}}
\end{equation}
where $r_{\rm e}$, $I_{\rm e}$, $n$ are the effective radius of the bulge, the intensity at $r_{\rm e}$ and the index parameter, respectively, while $b_n$ is a linear function of $n$ \citep[Eq.~6]{caon1993};
\begin{equation}
    \text{Disk:}\qquad I(r)=I_0\,{\rm e}^{-(r/r_{\rm d})}\qquad \text{\citep{freeman1970}}
\end{equation}
where $I_0$ and $r_{\rm d}$ are the central intensity and the scale parameter of the disk;
\begin{equation}
    \text{Broken disk:}\qquad I(r)=\begin{cases}
                                I_{0,1}\,{\rm e}^{(-r/r_{\rm d,1})} & r<r_{\rm br}\\
                                I_{0,2}\,{\rm e}^{(-r/r_{\rm d,2})} & r>r_{\rm br}
                            \end{cases}\qquad \text{\citep{pohlen2002}}
\end{equation}
where the two equations represent the inner and the outer behaviour of the disk SBP, both described by an exponential law with different scale parameters ($r_{\rm d,1}$, $r_{\rm d,2}$) and central intensities ($I_{0,1}$, $I_{0,2}$). The break radius $r_{\rm br}$ sets the change of the slope;
\begin{equation}
    \text{Cut-off disk:}\qquad I(r)=I_0\,{\rm e}^{-(r/r_{\rm d})-(r_{\rm c}/r)^3}\qquad \text{\citep{kormendy1977}}
\end{equation}
where the exponential disk is characterized by a inner drop for radii lower than the cut-off radius $r_{\rm c}$;
\begin{equation}
    \text{Lens:}\qquad I(r)=I_{0,{\rm l}}\,[1-(r/r_{\rm l})^2]\qquad \text{\citep{duval1983}}
\end{equation}
where $I_{\rm 0,l}$ and $r_{\rm l}$ are the central intensity and the size of the lens, respectively;
\begin{equation}
    \text{Bar:}\qquad I(r)=I_{0,{\rm b}}\,{[1-(r/a_{\rm b})^2]^{(n_{\rm b}+0.5)}}\qquad \text{\citep{ferrers1877}}
\end{equation}
where $I_{\rm 0,b}$, $a_{\rm b}$ and $n_{\rm b}$ are the central intensity, the size of the lens and the strength parameter of the bar, respectively.

Based on the visual inspection of the I band images and the extracted SBPs, for each galaxy we define a photometric model as the sum of the observed components. The best values of their free parameters are derived using the Python software package {\sc lmfit} \citep{newville2014} that implements a Non-Linear Least-Squares Minimization of the $\chi^2$
\begin{equation}
    \chi^2=\sum_i^N \frac{[I_{{\rm obs},i}-I_{{\rm mod},i}(\vec{p})]^2}{w_i^2},
\end{equation}
where $\vec{p}$ is the set of variables in the model $I_{\rm mod}$ and $I_{\rm obs}$ is the observed SBP. The fitting of the datapoints is achieved using a weighting factor $w_i$ that takes into account the errors of the isophote intensities and the fraction of the masked pixels along the isophotes.
Moreover, to reduce the seeing effect that levels off the central intensity we exclude the inner points within $0.5''$.

{ To test the quality of the fits, we have applied the Akaike information criterion (AIC, Aiake et al. 1973), which is an estimator that gives the relative quality of statistical models for a given set of data. 
Given a set of candidate models for the data, the preferred model is the one with the minimum AIC value. Thus, AIC rewards the goodness of a fit, but it also includes a penalty that is an increasing function of the number of estimated parameters. The penalty discourages overfitting, because increasing the number of parameters in the model almost always improves the goodness of the fit. We compare the AIC results of our decompositions to those obtained by performing a simple bulge+disk decomposition. As shown in Tab.\ref{tab:AIC}, the AIC value is always larger in the case of the bulge+disk decomposition, reassuring us about the choice of the model. We note that A3128\_B\_0248 is the only galaxy for which no third component exists.  Similar results are obtained also using the Bayesian information criterion (BIC).}

\begin{table}
\caption{AIC statistics for the adopted multicomponent fit (AIC$_{multi}$) and for a basic bulge+disk decomposition AIC$_{b+d}$.  \label{tab:AIC}}
\centering
\begin{small}
\begin{tabular}{l|c | c }
  \multicolumn{1}{c|}{id} &
  \multicolumn{1}{c|}{AIC$_{multi}$} &
  \multicolumn{1}{c}{AIC$_{b+d}$}\\
\hline
  A500\_22\_184 & -183.23 & -112.39 \\
  A500\_F\_0152 & -129.17 & -121.22\\
  A3158\_B\_0234 & -127.69 & -36.65\\
  A3376\_B\_0214 & -75.65 & 43.11 \\
  A3158\_11\_91 & -241.64 & -182.57\\
  A3128\_B\_0248 & -107.84 & -107.84\\
  A1069\_B\_0103 & 40.03 &56.61 \\
  A3158\_B\_0223 &-98.11 & -49.93 \\
\end{tabular}
\end{small}
\end{table}

\bibliography{references}{}

\begin{thebibliography}{}
\expandafter\ifx\csname natexlab\endcsname\relax\def\natexlab#1{#1}\fi
\providecommand{\url}[1]{\href{#1}{#1}}
\providecommand{\dodoi}[1]{doi:~\href{http://doi.org/#1}{\nolinkurl{#1}}}
\providecommand{\doeprint}[1]{\href{http://ascl.net/#1}{\nolinkurl{http://ascl.net/#1}}}
\providecommand{\doarXiv}[1]{\href{https://arxiv.org/abs/#1}{\nolinkurl{https://arxiv.org/abs/#1}}}

\bibitem[{Alatalo {et~al.}(2016)Alatalo, Cales, Rich, Appleton, Kewley, Lacy,
  Lanz, Medling, \& Nyland}]{Alatalo2016}
Alatalo, K., Cales, S.~L., Rich, J.~A., {et~al.} 2016, The Astrophysical
  Journal Supplement Series, 224, \dodoi{10.3847/0067-0049/224/2/38}

\bibitem[{Baldwin {et~al.}(1981)Baldwin, Phillips, \& Terlevich}]{Baldwin1981}
Baldwin, J.~A., Phillips, M.~M., \& Terlevich, R. 1981, {\textbackslash}pasp,
  93, 5, \dodoi{10.1086/130766}

\bibitem[{Balogh {et~al.}(2000)Balogh, Navarro, \&
  Morris}]{baloghnavarromorris00}
Balogh, M., Navarro, J., \& Morris, S. 2000, in KITP Conference: Galaxy
  Formation and Evolution

\bibitem[{Balogh {et~al.}(2004)Balogh, Baldry, Nichol, Miller, Bower, \&
  Glazebrook}]{Balogh2004}
Balogh, M.~L., Baldry, I.~K., Nichol, R., {et~al.} 2004, {\textbackslash}apjl,
  615, L101, \dodoi{10.1086/426079}

\bibitem[{Barro {et~al.}(2014)Barro, Faber, P{\'{e}}rez-Gonz{\'{a}}lez,
  Pacifici, Trump, Koo, Wuyts, Guo, Bell, Dekel, Porter, Primack, Ferguson,
  Ashby, Caputi, Ceverino, Croton, Fazio, Giavalisco, Hsu, Kocevski, Koekemoer,
  Kurczynski, Kollipara, Lee, McIntosh, McGrath, Moody, Somerville, Papovich,
  Salvato, Santini, Tal, van~der Wel, Williams, Willner, \&
  Zolotov}]{Barro2014}
Barro, G., Faber, S.~M., P{\'{e}}rez-Gonz{\'{a}}lez, P.~G., {et~al.} 2014, The
  Astrophysical Journal, 791, 52, \dodoi{10.1088/0004-637X/791/1/52}

\bibitem[{Bekki(1999)}]{Bekki1999}
Bekki, K. 1999, {\textbackslash}apjl, 510, L15, \dodoi{10.1086/311796}

\bibitem[{Bekki(2009)}]{Bekki2009}
---. 2009, {\textbackslash}mnras, 399, 2221,
  \dodoi{10.1111/j.1365-2966.2009.15431.x}

\bibitem[{Bekki(2014)}]{Bekki2014}
---. 2014, {\textbackslash}mnras, 438, 444, \dodoi{10.1093/mnras/stt2216}

\bibitem[{Bekki {et~al.}(2002)Bekki, Couch, \& Shioya}]{Bekki2002}
Bekki, K., Couch, W.~J., \& Shioya, Y. 2002, The Astrophysical Journal, 577,
  651, \dodoi{10.1086/342221}

\bibitem[{Bekki {et~al.}(2005)Bekki, Couch, Shioya, \& Vazdekis}]{Bekki2005}
Bekki, K., Couch, W.~J., Shioya, Y., \& Vazdekis, A. 2005.
\newblock \url{http://articles.adsabs.harvard.edu/pdf/2005MNRAS.359..949B}

\bibitem[{Bellhouse {et~al.}(2017)Bellhouse, Jaff{\'{e}}, Hau, McGee,
  Poggianti, Moretti, Gullieuszik, Bettoni, Fasano, D'Onofrio, Fritz, Omizzolo,
  Sheen, \& Vulcani}]{Bellhouse2017}
Bellhouse, C., Jaff{\'{e}}, Y., Hau, G., {et~al.} 2017, Astrophysical Journal,
  844, \dodoi{10.3847/1538-4357/aa7875}

\bibitem[{Bellhouse {et~al.}(2019)Bellhouse, Jaffe, McGee, Poggianti, Smith,
  Tonnesen, Fritz, Hau, Gullieuszik, Vulcani, Fasano, Moretti, George, Bettoni,
  D'Onofrio, Omizzolo, \& Sheen}]{Bellhouse2019}
Bellhouse, C., Jaffe, Y.~L., McGee, S.~L., {et~al.} 2019, Monthly Notices of
  the Royal Astronomical Society, Volume 485, Issue 1, p.1157-1170, 485, 1157,
  \dodoi{10.1093/mnras/stz460}

\bibitem[{Biviano {et~al.}(2017)Biviano, Moretti, Paccagnella, Poggianti,
  Bettoni, Gullieuszik, Vulcani, Fasano, D'onofrio, Fritz, \&
  Cava}]{Biviano2017}
Biviano, A., Moretti, A., Paccagnella, A., {et~al.} 2017, Astronomy and
  Astrophysics, 607, \dodoi{10.1051/0004-6361/201731289}

\bibitem[{Blake {et~al.}(2004)Blake, Pracy, Couch, Bekki, Lewis, Glazebrook,
  Baldry, Baugh, Bland-Hawthorn, Bridges, Cannon, Cole, Colless, Collins,
  Dalton, De~Propris, Driver, Efstathiou, Ellis, Frenk, Jackson, Lahav,
  Lumsden, Maddox, Madgwick, Norberg, Peacock, Peterson, Sutherland, \&
  Taylor}]{Blake2004}
Blake, C., Pracy, M.~B., Couch, W.~J., {et~al.} 2004, Mon. Not. R. Astron. Soc,
  355, 713, \dodoi{10.1111/j.1365-2966.2004.08351.x}

\bibitem[{Boselli {et~al.}(2014)Boselli, Cortese, Boquien, Boissier, Catinella,
  Gavazzi, Lagos, \& Saintonge}]{Boselli2014}
Boselli, A., Cortese, L., Boquien, M., {et~al.} 2014, Astronomy {\&}
  Astrophysics, 564, A67, \dodoi{10.1051/0004-6361/201322313}

\bibitem[{Boselli \& Gavazzi(2006)}]{Boselli2006}
Boselli, A., \& Gavazzi, G. 2006, {\textbackslash}pasp, 118, 517,
  \dodoi{10.1086/500691}

\bibitem[{Bournaud {et~al.}(2008)Bournaud, Duc, \& Emsellem}]{Bournaud2008}
Bournaud, F., Duc, P.-A., \& Emsellem, E. 2008, {\textbackslash}mnras, 389, L8,
  \dodoi{10.1111/j.1745-3933.2008.00511.x}

\bibitem[{Bressan {et~al.}(2012)Bressan, Marigo, Girardi, Salasnich, Dal~Cero,
  Rubele, \& Nanni}]{Bressan2012}
Bressan, A., Marigo, P., Girardi, L., {et~al.} 2012, {\textbackslash}mnras,
  427, 127, \dodoi{10.1111/j.1365-2966.2012.21948.x}

\bibitem[{Bryant {et~al.}(2015)Bryant, Owers, Robotham, Croom, Driver,
  Drinkwater, Lorente, Cortese, Scott, Colless, Schaefer, Taylor,
  Konstantopoulos, Allen, Baldry, Barnes, Bauer, Bland-Hawthorn, Bloom, Brooks,
  Brough, Cecil, Couch, Croton, Davies, Ellis, Fogarty, Foster, Glazebrook,
  Goodwin, Green, Gunawardhana, Hampton, Ho, Hopkins, Kewley, Lawrence,
  Leon-Saval, Leslie, McElroy, Lewis, Liske, L{\'{o}}pez-S{\'{a}}nchez,
  Mahajan, Medling, Metcalfe, Meyer, Mould, Obreschkow, O'Toole, Pracy,
  Richards, Shanks, Sharp, Sweet, Thomas, Tonini, \& Walcher}]{Bryant2015}
Bryant, J.~J., Owers, M.~S., Robotham, A.~S.~G., {et~al.} 2015,
  {\textbackslash}mnras, 447, 2857, \dodoi{10.1093/mnras/stu2635}

\bibitem[{Bundy {et~al.}(2015)Bundy, Bershady, Law, Yan, Drory, MacDonald,
  Wake, Cherinka, S{\'{a}}nchez-Gallego, Weijmans, Thomas, Tremonti, Masters,
  Coccato, Diamond-Stanic, Arag{\'{o}}n-Salamanca, Avila-Reese, Badenes,
  Falc{\'{o}}n-Barroso, Belfiore, Bizyaev, Blanc, Bland-Hawthorn, Blanton,
  Brownstein, Byler, Cappellari, Conroy, Dutton, Emsellem, Etherington,
  Frinchaboy, Fu, Gunn, Harding, Johnston, Kauffmann, Kinemuchi, Klaene,
  Knapen, Leauthaud, Li, Lin, Maiolino, Malanushenko, Malanushenko, Mao,
  Maraston, McDermid, Merrifield, Nichol, Oravetz, Pan, Parejko, Sanchez,
  Schlegel, Simmons, Steele, Steinmetz, Thanjavur, Thompson, Tinker, van~den
  Bosch, Westfall, Wilkinson, Wright, Xiao, \& Zhang}]{Bundy2015}
Bundy, K., Bershady, M.~A., Law, D.~R., {et~al.} 2015, {\textbackslash}apj,
  798, 7, \dodoi{10.1088/0004-637X/798/1/7}

\bibitem[{Byrd \& Valtonen(1990)}]{Byrd1990}
Byrd, G., \& Valtonen, M. 1990, {\textbackslash}apj, 350, 89,
  \dodoi{10.1086/168362}

\bibitem[{Caldwell {et~al.}(1996)Caldwell, Whipple, Rose, Franx, \&
  Leonardi}]{Caldwell1996}
Caldwell, N., Whipple, F.~L., Rose, J.~A., Franx, M., \& Leonardi, A.~J. 1996,
  {STARBURST COMA CLUSTER GALAXIES}, Tech. Rep.~1.
\newblock \url{http://articles.adsabs.harvard.edu/pdf/1996AJ....111...78C}

\bibitem[{Caon {et~al.}(1993)Caon, Capaccioli, \& D'Onofrio}]{caon1993}
Caon, N., Capaccioli, M., \& D'Onofrio, M. 1993, {\textbackslash}mnras, 265,
  1013, \dodoi{10.1093/mnras/265.4.1013}

\bibitem[{Cappellari \& Copin(2003)}]{Cappellari2003}
Cappellari, M., \& Copin, Y. 2003, {\textbackslash}mnras, 342, 345,
  \dodoi{10.1046/j.1365-8711.2003.06541.x}

\bibitem[{Cappellari \& Emsellem(2004)}]{Cappellari2004}
Cappellari, M., \& Emsellem, E. 2004, {\textbackslash}pasp, 116, 138,
  \dodoi{10.1086/381875}

\bibitem[{Cardelli {et~al.}(1989)Cardelli, Clayton, \& Mathis}]{Cardelli1989}
Cardelli, J.~A., Clayton, G.~C., \& Mathis, J.~S. 1989, {\textbackslash}apj,
  345, 245, \dodoi{10.1086/167900}

\bibitem[{Chabrier(2003)}]{Chabrier2003}
Chabrier, G. 2003, {Galactic Stellar and Substellar Initial Mass Function 1},
  Tech. rep.
\newblock \url{https://iopscience.iop.org/article/10.1086/376392/pdf}

\bibitem[{Chen {et~al.}(2019)Chen, Shi, Wild, Tremonti, Rowlands, Bizyaev, Yan,
  Lin, \& Riffel}]{Chen2019}
Chen, Y.-M., Shi, Y., Wild, V., {et~al.} 2019.
\newblock \url{http://arxiv.org/abs/1909.01658}

\bibitem[{Chilingarian {et~al.}(2009)Chilingarian, De~Rijcke, \&
  Buyle}]{Chilingarian2009}
Chilingarian, I.~V., De~Rijcke, S., \& Buyle, P. 2009, The Astrophysical
  Journal, 697, L111, \dodoi{10.1088/0004-637X/697/2/L111}

\bibitem[{Couch \& Sharples(1987)}]{couch87}
Couch, W.~J., \& Sharples, R.~M. 1987, {\textbackslash}mnras, 229, 423

\bibitem[{Davis {et~al.}(2019)Davis, van~de Voort, Rowlands, McAlpine, Wild, \&
  Crain}]{Davis2019}
Davis, T.~A., van~de Voort, F., Rowlands, K., {et~al.} 2019, Monthly Notices of
  the Royal Astronomical Society, 484, 2447, \dodoi{10.1093/mnras/stz180}

\bibitem[{De~Lucia {et~al.}(2012)De~Lucia, Weinmann, Poggianti,
  Arag{\'{o}}n-Salamanca, \& Zaritsky}]{DeLucia2012}
De~Lucia, G., Weinmann, S., Poggianti, B.~M., Arag{\'{o}}n-Salamanca, A., \&
  Zaritsky, D. 2012, {\textbackslash}mnras, 423, 1277,
  \dodoi{10.1111/j.1365-2966.2012.20983.x}

\bibitem[{Dressler \& Gunn(1982)}]{DresslerGunn1982}
Dressler, A., \& Gunn, J.~E. 1982, The Astrophysical Journal, 263, 533

\bibitem[{Dressler \& Gunn(1983)}]{dressler83}
---. 1983, {\textbackslash}apj, 270, 7, \dodoi{10.1086/161093}

\bibitem[{Dressler \& Gunn(1992)}]{DresslerGunn1992}
---. 1992, {SPECTROSCOPY OF GALAXIES IN DISTANT CLUSTERS. IV. A CATALOG OF
  PHOTOMETRY AND SPECTROSCOPY FOR GALAXIES IN SEVEN CLUSTERS WITH 0.35 < z <
  0.55}, Tech. rep.
\newblock \url{http://articles.adsabs.harvard.edu/pdf/1992ApJS...78....1D}

\bibitem[{Duval \& Athanassoula(1983)}]{duval1983}
Duval, M.~F., \& Athanassoula, E. 1983, {\textbackslash}aap, 121, 297

\bibitem[{Emsellem {et~al.}(2011)Emsellem, Cappellari, Krajnovi{\'{c}},
  Alatalo, Blitz, Bois, Bournaud, Bureau, Davies, Davis, de~Zeeuw, Khochfar,
  Kuntschner, Lablanche, McDermid, Morganti, Naab, Oosterloo, Sarzi, Scott,
  Serra, van~de Ven, Weijmans, \& Young}]{Emsellem2011}
Emsellem, E., Cappellari, M., Krajnovi{\'{c}}, D., {et~al.} 2011, Monthly
  Notices of the Royal Astronomical Society, 414, 888,
  \dodoi{10.1111/j.1365-2966.2011.18496.x}

\bibitem[{Fasano {et~al.}(2012)Fasano, Vanzella, Dressler, Poggianti, Moles,
  Bettoni, Valentinuzzi, Moretti, D'Onofrio, Varela, Couch, Kj{\ae}rgaard,
  Fritz, Omizzolo, Cava, Kjaergaard, Fritz, Omizzolo, \& Cava}]{Fasano2012}
Fasano, G., Vanzella, E., Dressler, A., {et~al.} 2012, {\textbackslash}mnras,
  420, 926, \dodoi{10.1111/j.1365-2966.2011.19798.x}

\bibitem[{Ferland {et~al.}(2013)Ferland, Porter, van Hoof, Williams, Abel,
  Lykins, Shaw, Henney, \& Stancil}]{Ferland2013}
Ferland, G.~J., Porter, R.~L., van Hoof, P.~A.~M., {et~al.} 2013,
  {\textbackslash}rmxaa, 49, 137

\bibitem[{Ferrers(1877)}]{ferrers1877}
Ferrers, N.~M. 1877, QJ Pure Appl. Math, 14, 1

\bibitem[{Freeman(1970)}]{freeman1970}
Freeman, K.~C. 1970, {\textbackslash}apj, 160, 811, \dodoi{10.1086/150474}

\bibitem[{French {et~al.}(2016)French, Arcavi, \& Zabludoff}]{French2016}
French, K.~D., Arcavi, I., \& Zabludoff, A. 2016, The Astrophysical Journal,
  818, L21, \dodoi{10.3847/2041-8205/818/1/L21}

\bibitem[{French {et~al.}(2015)French, Yang, Zabludoff, Narayanan, Shirley,
  Walter, Smith, \& Tremonti}]{French2015}
French, K.~D., Yang, Y., Zabludoff, A., {et~al.} 2015, The Astrophysical
  Journal, 801, 1, \dodoi{10.1088/0004-637X/801/1/1}

\bibitem[{Fritz {et~al.}(2007)Fritz, Poggianti, Bettoni, Cava, Couch,
  D'Onofrio, Dressler, Fasano, Kj{\ae}rgaard, Moles, \& Varela}]{Fritz2007}
Fritz, J., Poggianti, B.~M., Bettoni, D., {et~al.} 2007, {\textbackslash}aap,
  470, 137, \dodoi{10.1051/0004-6361:20077097}

\bibitem[{Fritz {et~al.}(2011)Fritz, Poggianti, Cava, Valentinuzzi, Moretti,
  Bettoni, Bressan, Couch, D'Onofrio, Dressler, Fasano, Kj{\ae}rgaard, Moles,
  Omizzolo, \& Varela}]{Fritz2011}
Fritz, J., Poggianti, B.~M., Cava, A., {et~al.} 2011, {\textbackslash}aap, 526,
  A45+, \dodoi{10.1051/0004-6361/201015214}

\bibitem[{Fritz {et~al.}(2017)Fritz, Moretti, Gullieuszik, Poggianti, Bruzual,
  Vulcani, Nicastro, Jaff{\'{e}}, Cervantes~Sodi, Bettoni, Biviano, Fasano,
  Charlot, Bellhouse, \& Hau}]{Fritz2017}
Fritz, J., Moretti, A., Gullieuszik, M., {et~al.} 2017, Astrophysical Journal,
  848, \dodoi{10.3847/1538-4357/aa8f51}

\bibitem[{George {et~al.}(2018)George, Poggianti, Gullieuszik, Fasano,
  Bellhouse, Postma, Moretti, Jaff{\'{e}}, Vulcani, Bettoni, Fritz,
  C{\^{o}}t{\'{e}}, Ghosh, Hutchings, Mohan, Sreekumar, Stalin, Subramaniam, \&
  Tandon}]{George2018}
George, K., Poggianti, B., Gullieuszik, M., {et~al.} 2018, Monthly Notices of
  the Royal Astronomical Society, 479, \dodoi{10.1093/mnras/sty1452}

\bibitem[{George {et~al.}(2019)George, Poggianti, Bellhouse, Radovich, Fritz,
  Paladino, Bettoni, Jaff{\'{e}}, Moretti, Gullieuszik, Vulcani, Fasano,
  Stalin, Subramaniam, \& Tandon}]{George2019}
George, K., Poggianti, B.~M., Bellhouse, C., {et~al.} 2019, Monthly Notices of
  the Royal Astronomical Society, 487, 3102, \dodoi{10.1093/mnras/stz1443}

\bibitem[{Ghigna {et~al.}(1998)Ghigna, Moore, Governato, Lake, Quinn, \&
  Stadel}]{Ghigna1998}
Ghigna, S., Moore, B., Governato, F., {et~al.} 1998, Monthly Notices of the
  Royal Astronomical Society, 300, 146

\bibitem[{Gladders {et~al.}(2013)Gladders, Oemler, Dressler, Poggianti,
  Vulcani, \& Abramson}]{Gladders2013}
Gladders, M.~D., Oemler, A., Dressler, A., {et~al.} 2013, Astrophysical
  Journal, 770, \dodoi{10.1088/0004-637X/770/1/64}

\bibitem[{Goto(2005)}]{Goto2005}
Goto, T. 2005, Mon. Not. R. Astron. Soc, 357, 937, \dodoi{10.111}

\bibitem[{Graham {et~al.}(2019)Graham, Cappellari, Bershady, \&
  Drory}]{Graham2019}
Graham, M.~T., Cappellari, M., Bershady, M.~A., \& Drory, N. 2019.
\newblock \url{http://arxiv.org/abs/1911.06103}

\bibitem[{Gullieuszik {et~al.}(2015)Gullieuszik, Poggianti, Fasano, Zaggia,
  Paccagnella, Moretti, Bettoni, Couch, Vulcani, Fritz, Omizzolo, Baruffolo,
  Schipani, Capaccioli, \& Varela}]{Gullieuszik2015}
Gullieuszik, M., Poggianti, B., Fasano, G., {et~al.} 2015, A{\&}A, 581, 41,
  \dodoi{10.1051/0004-6361/201526061}

\bibitem[{Gullieuszik {et~al.}(2017)Gullieuszik, Poggianti, Moretti, Fritz,
  Jaff{\'{e}}, Hau, Bischko, Bellhouse, Bettoni, Fasano, Vulcani, D'Onofrio, \&
  Biviano}]{Gullieuszik2017}
Gullieuszik, M., Poggianti, B., Moretti, A., {et~al.} 2017, Astrophysical
  Journal, 846, \dodoi{10.3847/1538-4357/aa8322}

\bibitem[{Gunn \& Gott(1972)}]{Gunn1972}
Gunn, J.~E., \& Gott, J.~R. 1972, {\textbackslash}apj, 176, 1,
  \dodoi{10.1086/151605}

\bibitem[{Haines {et~al.}(2015)Haines, Pereira, Smith, Egami, Babul,
  Finoguenov, Ziparo, McGee, Rawle, Okabe, \& Moran}]{Haines2015}
Haines, C.~P., Pereira, M.~J., Smith, G.~P., {et~al.} 2015,
  {\textbackslash}apj, 806, 101, \dodoi{10.1088/0004-637X/806/1/101}

\bibitem[{Haines {et~al.}(2018)Haines, Finoguenov, Smith, Babul, Egami,
  Mazzotta, Okabe, Pereira, Bianconi, McGee, Ziparo, Campusano, \&
  Loyola}]{Haines2018}
Haines, C.~P., Finoguenov, A., Smith, G.~P., {et~al.} 2018, Monthly Notices of
  the Royal Astronomical Society, 477, 4931, \dodoi{10.1093/mnras/sty651}

\bibitem[{Hopkins {et~al.}(2009)Hopkins, Cox, Dutta, Hernquist, Kormendy, \&
  Lauer}]{Hopkins2009}
Hopkins, P.~F., Cox, T.~J., Dutta, S.~N., {et~al.} 2009, The Astrophysical
  Journal Supplement Series, 181, 135, \dodoi{10.1088/0067-0049/181/1/135}

\bibitem[{Hopkins {et~al.}(2006)Hopkins, Hernquist, Cox, Di~Matteo, Robertson,
  \& Springel}]{Hopkins2006}
Hopkins, P.~F., Hernquist, L., Cox, T.~J., {et~al.} 2006, The Astrophysical
  Journal Supplement Series, 163, 1, \dodoi{10.1086/499298}

\bibitem[{Jaff{\'{e}} {et~al.}(2018)Jaff{\'{e}}, Poggianti, Moretti,
  Gullieuszik, Smith, Vulcani, Fasano, Fritz, Tonnesen, Bettoni, Hau, Biviano,
  Bellhouse, \& McGee}]{Jaffe2018}
Jaff{\'{e}}, Y., Poggianti, B., Moretti, A., {et~al.} 2018, Monthly Notices of
  the Royal Astronomical Society, 476, \dodoi{10.1093/mnras/sty500}

\bibitem[{Jaff{\'{e}} {et~al.}(2015)Jaff{\'{e}}, Smith, Candlish, Poggianti,
  Sheen, \& Verheijen}]{Jaffe2015}
Jaff{\'{e}}, Y.~L., Smith, R., Candlish, G.~N., {et~al.} 2015,
  {\textbackslash}mnras, 448, 1715, \dodoi{10.1093/mnras/stv100}

\bibitem[{Jedrzejewski(1987)}]{Jedrzejewski1987}
Jedrzejewski, R.~I. 1987, {\textbackslash}mnras, 226, 747,
  \dodoi{10.1093/mnras/226.4.747}

\bibitem[{Klitsch {et~al.}(2017)Klitsch, Zwaan, Kuntschner, Couch, Pracy, \&
  Owers}]{Klitsch2017}
Klitsch, A., Zwaan, M.~A., Kuntschner, H., {et~al.} 2017, A{\&}A, 600, 80,
  \dodoi{10.1051/0004-6361/201527922}

\bibitem[{Kormendy(1977)}]{kormendy1977}
Kormendy, J. 1977, {\textbackslash}apj, 217, 406, \dodoi{10.1086/155589}

\bibitem[{Kronberger {et~al.}(2008)Kronberger, Kapferer, Ferrari,
  Unterguggenberger, \& Schindler}]{Kronberger2008}
Kronberger, T., Kapferer, W., Ferrari, C., Unterguggenberger, S., \& Schindler,
  S. 2008, {\textbackslash}aap, 481, 337, \dodoi{10.1051/0004-6361:20078904}

\bibitem[{Larson {et~al.}(1980)Larson, Tinsley, \& Caldwell}]{Larson1980}
Larson, R.~B., Tinsley, B.~M., \& Caldwell, C.~N. 1980, {\textbackslash}apj,
  237, 692, \dodoi{10.1086/157917}

\bibitem[{Lee {et~al.}(2017)Lee, Chung, Tonnesen, Kenney, Wong, Vollmer,
  Petitpas, Crowl, \& van Gorkom}]{Lee2017}
Lee, B., Chung, A., Tonnesen, S., {et~al.} 2017, Monthly Notices of the Royal
  Astronomical Society, 466, 1382, \dodoi{10.1093/mnras/stw3162}

\bibitem[{Martin {et~al.}(2007)Martin, Wyder, Schiminovich, Barlow, Forster,
  Friedman, Morrissey, Neff, Seibert, Small, Welsh, Bianchi, Donas, Heckman,
  Lee, Madore, Milliard, Rich, Szalay, \& Yi}]{Martin2007}
Martin, D.~C., Wyder, T.~K., Schiminovich, D., {et~al.} 2007, The Astrophysical
  Journal Supplement Series, 173, 342, \dodoi{10.1086/516639}

\bibitem[{McGee {et~al.}(2009)McGee, Balogh, Bower, Font, \&
  McCarthy}]{McGee2009}
McGee, S.~L., Balogh, M.~L., Bower, R.~G., Font, A.~S., \& McCarthy, I.~G.
  2009, {\textbackslash}mnras, 400, 937,
  \dodoi{10.1111/j.1365-2966.2009.15507.x}

\bibitem[{Moore {et~al.}(1996)Moore, Katz, Lake, Dressler, \&
  Oemler}]{Moore1996}
Moore, B., Katz, N., Lake, G., Dressler, A., \& Oemler, A. 1996,
  {\textbackslash}nat, 379, 613, \dodoi{10.1038/379613a0}

\bibitem[{Moretti {et~al.}(2017)Moretti, Gullieuszik, Poggianti, Paccagnella,
  Couch, Vulcani, Bettoni, Fritz, Cava, Fasano, \& Omizzolo}]{Moretti2017}
Moretti, A., Gullieuszik, M., Poggianti, B., {et~al.} 2017, A{\&}A, 599, 81,
  \dodoi{10.1051/0004-6361/201630030}

\bibitem[{Moretti {et~al.}(2018)Moretti, Poggianti, Gullieuszik, Mapelli,
  Jaff{\'{e}}, Fritz, Biviano, Fasano, Bettoni, Vulcani, \&
  D'Onofrio}]{Moretti2018}
Moretti, A., Poggianti, B.~M., Gullieuszik, M., {et~al.} 2018,
  {\textbackslash}mnras, 475, 4055, \dodoi{10.1093/mnras/sty085}

\bibitem[{Newville {et~al.}(2014)Newville, Stensitzki, Allen, \&
  Ingargiola}]{newville2014}
Newville, M., Stensitzki, T., Allen, D.~B., \& Ingargiola, A. 2014, {LMFIT:
  Non-Linear Least-Square Minimization and Curve-Fitting for Python},
  \dodoi{10.5281/zenodo.11813}

\bibitem[{Norton {et~al.}(2001)Norton, Gebhardt, Zabludoff, \&
  Zaritsky}]{Norton2001}
Norton, S.~A., Gebhardt, K., Zabludoff, A.~I., \& Zaritsky, D. 2001, The
  Astrophysical Journal, 557, 150, \dodoi{10.1086/321668}

\bibitem[{Nulsen(1982)}]{Nulsen1982}
Nulsen, P.~E.~J. 1982, {\textbackslash}mnras, 198, 1007

\bibitem[{Oman {et~al.}(2013)Oman, Hudson, \& Behroozi}]{Oman2013}
Oman, K.~A., Hudson, M.~J., \& Behroozi, P.~S. 2013, {\textbackslash}mnras,
  431, 2307, \dodoi{10.1093/mnras/stt328}

\bibitem[{Owers {et~al.}(2019)Owers, Hudson, Oman, Bland-Hawthorn, Brough,
  Bryant, Cortese, Couch, Croom, Sande, Federrath, Groves, Hopkins, Lawrence,
  Lorente, McDermid, Medling, Richards, Scott, Taranu, Welker, \&
  Yi}]{Owers2019}
Owers, M.~S., Hudson, M.~J., Oman, K.~A., {et~al.} 2019, The Astrophysical
  Journal, 873, 52, \dodoi{10.3847/1538-4357/ab0201}

\bibitem[{Paccagnella {et~al.}(2019)Paccagnella, Vulcani, Poggianti, Moretti,
  Fritz, Gullieuszik, \& Fasano}]{Paccagnella2019}
Paccagnella, A., Vulcani, B., Poggianti, B.~M., {et~al.} 2019, Monthly Notices
  of the Royal Astronomical Society, 482, 881, \dodoi{10.1093/mnras/sty2728}

\bibitem[{Paccagnella {et~al.}(2017)Paccagnella, Vulcani, Poggianti, Fritz,
  Fasano, Moretti, Jaff{\'{e}}, Biviano, Gullieuszik, Bettoni, Cava, Couch, \&
  D'onofrio}]{Paccagnella2017}
---. 2017, The Astrophysical Journal, 838, 148,
  \dodoi{10.3847/1538-4357/aa64d7}

\bibitem[{Pattarakijwanich {et~al.}(2016)Pattarakijwanich, Strauss, Ho, \&
  Ross}]{Pattarakijwanich2016}
Pattarakijwanich, P., Strauss, M.~A., Ho, S., \& Ross, N.~P. 2016, The
  Astrophysical Journal, 833, 19, \dodoi{10.3847/0004-637X/833/1/19}

\bibitem[{Pawlik {et~al.}(2019)Pawlik, McAlpine, Trayford, Wild, Bower, Crain,
  Schaller, \& Schaye}]{Pawlik2019}
Pawlik, M.~M., McAlpine, S., Trayford, J.~W., {et~al.} 2019, Nature Astronomy,
  3, 440, \dodoi{10.1038/s41550-019-0725-z}

\bibitem[{Poggianti {et~al.}(1999)Poggianti, Smail, Dressler, Couch, Barger,
  Butcher, Ellis, \& Oemler~Jr.}]{poggianti99}
Poggianti, B.~M., Smail, I., Dressler, A., {et~al.} 1999, {\textbackslash}apj,
  518, 576, \dodoi{10.1086/307322}

\bibitem[{Poggianti {et~al.}(2009)Poggianti, Arag{\'{o}}n-Salamanca, Zaritsky,
  De~Lucia, Milvang-Jensen, Desai, Jablonka, Halliday, Rudnick, Varela,
  Bamford, Best, Clowe, Noll, Saglia, Pell{\'{o}}, Simard, von~der Linden, \&
  White}]{Poggianti2009b}
Poggianti, B.~M., Arag{\'{o}}n-Salamanca, A., Zaritsky, D., {et~al.} 2009, The
  Astrophysical Journal, 693, 112, \dodoi{10.1088/0004-637X/693/1/112}

\bibitem[{Poggianti {et~al.}(2017)Poggianti, Moretti, Gullieuszik, Fritz,
  Jaff{\'{e}}, Bettoni, Fasano, Bellhouse, Hau, Vulcani, Biviano, Omizzolo,
  Paccagnella, D'onofrio, Cava, Sheen, Couch, \& Owers}]{Poggianti2017}
Poggianti, B.~M., Moretti, A., Gullieuszik, M., {et~al.} 2017,
  \dodoi{10.3847/1538-4357/aa78ed}

\bibitem[{Poggianti {et~al.}(2019)Poggianti, Ignesti, Gitti, Wolter, Brighenti,
  Biviano, George, Vulcani, Gullieuszik, Moretti, Paladino, Bettoni,
  Franchetto, Jaff{\'{e}}, Radovich, Roediger, Tomi{\v{c}}i´c, Tonnesen,
  Bellhouse, Fritz, \& Omizzolo}]{Poggianti2019b}
Poggianti, B.~M., Ignesti, A., Gitti, M., {et~al.} 2019, {GASP XXIII: A
  jellyfish galaxy as an astrophysical laboratory of the baryonic cycle}, Tech.
  rep.
\newblock \url{https://www.eso.org/public/news/eso1725/}

\bibitem[{Pohlen {et~al.}(2002)Pohlen, Dettmar, L{\"{u}}tticke, \&
  Aronica}]{pohlen2002}
Pohlen, M., Dettmar, R.~J., L{\"{u}}tticke, R., \& Aronica, G. 2002,
  {\textbackslash}aap, 392, 807, \dodoi{10.1051/0004-6361:20020994}

\bibitem[{Pracy {et~al.}(2005)Pracy, Couch, Blake, Bekki, Harrison, Colless,
  Kuntschner, \& De~Propris}]{Pracy2005}
Pracy, M.~B., Couch, W.~J., Blake, C., {et~al.} 2005, Mon. Not. R. Astron. Soc,
  35, 1421, \dodoi{10.111}

\bibitem[{Pracy {et~al.}(2009)Pracy, Kuntschner, Couch, Blake, Bekki, \&
  Briggs}]{Pracy2009}
Pracy, M.~B., Kuntschner, H., Couch, W.~J., {et~al.} 2009, Monthly Notices of
  the Royal Astronomical Society, 396, 1349,
  \dodoi{10.1111/j.1365-2966.2009.14836.x}

\bibitem[{Pracy {et~al.}(2012)Pracy, Owers, Couch, Kuntschner, Bekki, Briggs,
  Lah, \& Zwaan}]{Pracy2012}
Pracy, M.~B., Owers, M.~S., Couch, W.~J., {et~al.} 2012, Monthly Notices of the
  Royal Astronomical Society, 420, 2232,
  \dodoi{10.1111/j.1365-2966.2011.20188.x}

\bibitem[{Pracy {et~al.}(2013)Pracy, Croom, Sadler, Couch, Kuntschner, Bekki,
  Owers, Zwaan, Turner, \& Bergmann}]{Pracy2013}
Pracy, M.~B., Croom, S., Sadler, E., {et~al.} 2013, Monthly Notices of the
  Royal Astronomical Society, 432, 3131, \dodoi{10.1093/mnras/stt666}

\bibitem[{Rhee {et~al.}(2017)Rhee, Smith, Choi, Yi, Jaff{\'{e}}, Candlish, \&
  S{\'{a}}nchez-J{\'{a}}nssen}]{Rhee2017}
Rhee, J., Smith, R., Choi, H., {et~al.} 2017, {\textbackslash}apj, 843, 128,
  \dodoi{10.3847/1538-4357/aa6d6c}

\bibitem[{Roche {et~al.}(2015)Roche, Humphrey, Gomes, Papaderos, Lagos, \&
  S{\'{a}}nchez}]{Roche2015}
Roche, N., Humphrey, A., Gomes, J.~M., {et~al.} 2015, Monthly Notices of the
  Royal Astronomical Society, 453, 2350, \dodoi{10.1093/mnras/stv1669}

\bibitem[{Roediger \& Br{\"{u}}ggen(2006)}]{RoedigerBruggen2006}
Roediger, E., \& Br{\"{u}}ggen, M. 2006, {\textbackslash}mnras, 369, 567,
  \dodoi{10.1111/j.1365-2966.2006.10335.x}

\bibitem[{Rowlands {et~al.}(2015)Rowlands, Wild, Nesvadba, Sibthorpe, Mortier,
  Lehnert, \& da~Cunha}]{Rowlands2015}
Rowlands, K., Wild, V., Nesvadba, N., {et~al.} 2015, Monthly Notices of the
  Royal Astronomical Society, 448, 258, \dodoi{10.1093/mnras/stu2714}

\bibitem[{Rowlands {et~al.}(2018)Rowlands, Heckman, Wild, Zakamska,
  Rodriguez-Gomez, Barrera-Ballesteros, Lotz, Thilker, Andrews, Boquien,
  Brinkmann, Brownstein, Hwang, \& Smethurst}]{Rowlands2018}
Rowlands, K., Heckman, T., Wild, V., {et~al.} 2018, Monthly Notices of the
  Royal Astronomical Society, 480, 2544, \dodoi{10.1093/mnras/sty1916}

\bibitem[{S{\'{a}}nchez {et~al.}(2006)S{\'{a}}nchez, Baugh, Percival, Peacock,
  Padilla, Cole, Frenk, \& Norberg}]{sanchez06}
S{\'{a}}nchez, A.~G., Baugh, C.~M., Percival, W.~J., {et~al.} 2006,
  {\textbackslash}mnras, 366, 189, \dodoi{10.1111/j.1365-2966.2005.09833.x}

\bibitem[{Sanders {et~al.}(1988)Sanders, Soifer, Elias, Madore, Matthews,
  Neugebauer, \& Scoville}]{Sanders1988}
Sanders, D.~B., Soifer, B.~T., Elias, J.~H., {et~al.} 1988, The Astrophysical
  Journal, 325, 74, \dodoi{10.1086/165983}

\bibitem[{Schawinski {et~al.}(2014)Schawinski, Urry, Simmons, Fortson, Kaviraj,
  Keel, Lintott, Masters, Nichol, Sarzi, Skibba, Treister, Willett, Wong, \&
  Yi}]{Schawinski2014}
Schawinski, K., Urry, C.~M., Simmons, B.~D., {et~al.} 2014, Monthly Notices of
  the Royal Astronomical Society, 440, 889, \dodoi{10.1093/mnras/stu327}

\bibitem[{Sersic(1968)}]{sersic1968}
Sersic, J.~L. 1968, {Atlas de Galaxias Australes}

\bibitem[{Smith {et~al.}(2015)Smith, S{\'{a}}nchez-Janssen, Beasley, Candlish,
  Gibson, Puzia, Janz, Knebe, Aguerri, Lisker, Hensler, Fellhauer, Ferrarese,
  \& Yi}]{Smith2015}
Smith, R., S{\'{a}}nchez-Janssen, R., Beasley, M.~A., {et~al.} 2015,
  {\textbackslash}mnras, 454, 2502, \dodoi{10.1093/mnras/stv2082}

\bibitem[{Snyder {et~al.}(2011)Snyder, Hopkins, \& Hernquist}]{Snyder2011}
Snyder, G.~F., Hopkins, P.~F., \& Hernquist, L. 2011, The Astrophysical
  Journal, 728, L24, \dodoi{10.1088/2041-8205/728/1/L24}

\bibitem[{Socolovsky {et~al.}(2018)Socolovsky, Almaini, Hatch, Wild, Maltby,
  Hartley, \& Simpson}]{Socolovsky2018}
Socolovsky, M., Almaini, O., Hatch, N.~A., {et~al.} 2018, Monthly Notices of
  the Royal Astronomical Society, 476, 1242, \dodoi{10.1093/mnras/sty312}

\bibitem[{Swinbank {et~al.}(2012)Swinbank, Balogh, Bower, Zabludoff, Lucey,
  McGee, Miller, \& Nichol}]{Swinbank2012}
Swinbank, A.~M., Balogh, M.~L., Bower, R.~G., {et~al.} 2012, Monthly Notices of
  the Royal Astronomical Society, 420, 672,
  \dodoi{10.1111/j.1365-2966.2011.20082.x}

\bibitem[{Tran {et~al.}(2003)Tran, Franx, Illingworth, Kelson, \& van
  Dokkum}]{Tran2003}
Tran, K.~H., Franx, M., Illingworth, G., Kelson, D.~D., \& van Dokkum, P. 2003,
  The Astrophysical Journal, 599, 865, \dodoi{10.1086/379804}

\bibitem[{Tran {et~al.}(2007)Tran, Franx, Illingworth, van Dokkum, Kelson,
  Blakeslee, \& Postman}]{Tran2007}
Tran, K.~H., Franx, M., Illingworth, G.~D., {et~al.} 2007, The Astrophysical
  Journal, 661, 750, \dodoi{10.1086/513738}

\bibitem[{Tran {et~al.}(2004)Tran, Franx, Illingworth, van Dokkum, Kelson, \&
  Magee}]{Tran2004}
---. 2004, The Astrophysical Journal, 609, 683, \dodoi{10.1086/421237}

\bibitem[{Vulcani {et~al.}(2015)Vulcani, Poggianti, Fritz, Fasano, Moretti,
  Calvi, \& Paccagnella}]{Vulcani2015}
Vulcani, B., Poggianti, B.~M., Fritz, J., {et~al.} 2015, The Astrophysical
  Journal, 798, 52, \dodoi{10.1088/0004-637X/798/1/52}

\bibitem[{Vulcani {et~al.}(2012)Vulcani, Poggianti, Fasano, Desai, Dressler,
  Oemler, Calvi, D’Onofrio, \& Moretti}]{Vulcani2012}
Vulcani, B., Poggianti, B.~M., Fasano, G., {et~al.} 2012, Monthly Notices of
  the Royal Astronomical Society, 420, 1481,
  \dodoi{10.1111/j.1365-2966.2011.20135.x}

\bibitem[{Vulcani {et~al.}(2018{\natexlab{a}})Vulcani, Poggianti, Gullieuszik,
  Moretti, Tonnesen, Jaff{\'{e}}, Fritz, Fasano, \& Bettoni}]{Vulcani2018_L}
Vulcani, B., Poggianti, B.~M., Gullieuszik, M., {et~al.} 2018{\natexlab{a}},
  The Astrophysical Journal Letters, 866, L25, \dodoi{10.3847/2041-8213/aae68b}

\bibitem[{Vulcani {et~al.}(2018{\natexlab{b}})Vulcani, Poggianti, Jaff{\'{e}},
  Moretti, Fritz, Gullieuszik, Bettoni, Fasano, Tonnesen, \&
  McGee}]{Vulcani2018_g}
Vulcani, B., Poggianti, B., Jaff{\'{e}}, Y., {et~al.} 2018{\natexlab{b}},
  Monthly Notices of the Royal Astronomical Society, 480,
  \dodoi{10.1093/MNRAS/STY2095}

\bibitem[{Vulcani {et~al.}(2019)Vulcani, Poggianti, Moretti, Franchetto,
  Gullieuszik, Fritz, Bettoni, Tonnesen, Radovich, Jaff{\'{e}}, McGee,
  Bellhouse, \& Fasano}]{Vulcani2019b}
Vulcani, B., Poggianti, B.~M., Moretti, A., {et~al.} 2019, Monthly Notices of
  the Royal Astronomical Society, 1773, \dodoi{10.1093/mnras/stz1829}

\bibitem[{Wong {et~al.}(2012)Wong, Schawinski, Kaviraj, Masters, Nichol,
  Lintott, Keel, Darg, Bamford, Andreescu, Murray, Raddick, Szalay, Thomas, \&
  VandenBerg}]{Wong2012}
Wong, O.~I., Schawinski, K., Kaviraj, S., {et~al.} 2012, Monthly Notices of the
  Royal Astronomical Society, 420, 1684,
  \dodoi{10.1111/j.1365-2966.2011.20159.x}

\bibitem[{Wu {et~al.}(2014)Wu, Gal, Lemaux, Kocevski, Lubin, Rumbaugh, \&
  Squires}]{Wu2014}
Wu, P.-F., Gal, R.~R., Lemaux, B.~C., {et~al.} 2014, The Astrophysical Journal,
  792, 16, \dodoi{10.1088/0004-637X/792/1/16}

\bibitem[{Yagi \& Goto(2006)}]{Yagi2006}
Yagi, M., \& Goto, T. 2006, The Astronomical Journal, 131, 2050,
  \dodoi{10.1086/500660}

\bibitem[{Yang {et~al.}(2008)Yang, Mo, \& van~den Bosch}]{Yang2008}
Yang, X., Mo, H.~J., \& van~den Bosch, F.~C. 2008, The Astrophysical Journal,
  676, 248, \dodoi{10.1086/528954}

\bibitem[{Yang {et~al.}(2004)Yang, Zabludoff, Zaritsky, Lauer, \&
  Mihos}]{Yang2004}
Yang, Y., Zabludoff, A.~I., Zaritsky, D., Lauer, T.~R., \& Mihos, J.~C. 2004,
  The Astrophysical Journal, 607, 258, \dodoi{10.1086/383259}

\bibitem[{Yesuf {et~al.}(2014)Yesuf, Faber, Trump, Koo, Fang, Liu, Wild, \&
  Hayward}]{Yesuf2014}
Yesuf, H.~M., Faber, S.~M., Trump, J.~R., {et~al.} 2014, The Astrophysical
  Journal, 792, 84, \dodoi{10.1088/0004-637X/792/2/84}

\bibitem[{Yoon {et~al.}(2017)Yoon, Chung, Smith, \& Jaff{\'{e}}}]{Yoon2017}
Yoon, H., Chung, A., Smith, R., \& Jaff{\'{e}}, Y.~L. 2017,
  {\textbackslash}apj, 838, 81, \dodoi{10.3847/1538-4357/aa6579}

\bibitem[{Zabludoff {et~al.}(1996)Zabludoff, Zaritsky, Lin, Tucker, Hashimoto,
  Shectman, Oemler, \& Kirshner}]{zabludoff96}
Zabludoff, A.~I., Zaritsky, D., Lin, H., {et~al.} 1996, {\textbackslash}apj,
  466, 104, \dodoi{10.1086/177495}

\bibitem[{Zwaan {et~al.}(2013)Zwaan, Kuntschner, Pracy, \& Couch}]{Zwaan2013}
Zwaan, M.~A., Kuntschner, H., Pracy, M.~B., \& Couch, W.~J. 2013, Monthly
  Notices of the Royal Astronomical Society, 432, 492,
  \dodoi{10.1093/mnras/stt496}

\end{thebibliography}
\bibliographystyle{aasjournal}



\end{document}